\documentclass[a4paper,twoside]{article}
\pdfoutput=1 
\newcommand{\affil}[1]{$^{\rm #1}$}
\usepackage[authoryear]{natbib}
\bibpunct{(}{)}{;}{a}{}{,}
\usepackage{amssymb}
\usepackage{natbib, fancyheadings}
\usepackage{graphics, graphicx}
\usepackage{times}
\date{} %Please leave the date blank
\title{\large\bf\flushleft The completeness and reliability of threshold and false-discovery-rate source extraction algorithms for compact continuum sources.}
\author{\parbox{\textwidth}{\flushleft
\vspace{-0.5cm}
{\it Minh T. Huynh\affil{A,G}, Andrew Hopkins\affil{B,F}, Ray Norris\affil{C}, Paul Hancock\affil{D, F}, Tara Murphy\affil{D,E, F}, Russell Jurek\affil{C}, and Matthew Whiting\affil{C}}\\
\vspace{0.4cm}
{\small \affil{A}\,International Centre for Radio Astronomy Research, M468, University of Western
Australia, Crawley, WA 6009, Australia}\\
{\small \affil{B}\,Australian Astronomical Observatory, P.O. Box 296, Epping NSW 1710, Australia}\\
{\small \affil{C}\,CSIRO Astronomy \& Space Sciences, Australia Telescope National Facility, P.O. Box 76, Epping NSW 1710, Australia}\\
{\small \affil{D}\,Sydney Institute for Astronomy, School of Physics, The University of Sydney, NSW 2006, Australia}\\
{\small \affil{E}\,School of Information Technologies, The University of Sydney, NSW 2006, Australia }\\
{\small \affil{F}\, ARC Centre of Excellence for All-sky Astrophysics (CAASTRO)}\\
{\small \affil{G}\,Email: minh.huynh@uwa.edu.au}}}
\begin{document}
\twocolumn[
\begin{changemargin}{.8cm}{.5cm}
\begin{minipage}{.9\textwidth}
\vspace{-1cm}
\maketitle
%
%
%%%%%%%%%%%%%     ABSTRACT    %%%%%%%%%%%%%
%Abstract of no more than 200 words here.
\small{\bf 
The process of determining the number and characteristics of sources
in astronomical images is so fundamental to a large range of
astronomical problems that it is perhaps surprising that no standard procedure has ever been defined
that has well understood properties with a high degree of statistical rigour
on completeness and reliability.
There are now a large number of commonly used software tools for
accomplishing this task, typically with different tools being used for images acquired using different
technologies (e.g., radio telescope images as compared to optical/NIR images).
Despite this, there have been relatively few quantitative analyses of the robustness or reliability of
individual tools, or the details of the techniques they implement. We have an opportunity to
redress this omission in the context of surveys planned with the Australian Square Kilometre
Array Pathfinder (ASKAP). The Evolutionary Map of the Universe (EMU) survey with ASKAP, a
continuum survey of the Southern Hemisphere up to declination $+30^{\circ}$, aims to
utilise an automated source identification and measurement approach that is demonstrably
optimal, to maximise the reliability, utility and robustness of the resulting radio source catalogues.
A key stage in source extraction methods is the background estimation (background level
and noise level) and the choice of a threshold high enough to reject false
sources yet not so high that the catalogues are significantly incomplete. It is important to
characterise the performance of such algorithm(s) that may potentially be used within
the EMU and ASKAP source extraction pipeline. In this analysis we present results
from testing such algorithms as implemented in the SExtractor, Selavy (Duchamp), and {\sc sfind} tools
on simulated data. In particular the effects of background estimation, threshold and false-discovery
rate settings are explored. For parameters that give similar completeness, the false-discovery rate method employed 
by {\sc sfind} results in a more reliable catalogue compared to the peak threshold methods of SExtractor and Selavy.}

%%%%%%%%%%%%%     KEYWORDS    %%%%%%%%%%%%%
\medskip{\bf Keywords:}  methods: data analysis --- radio continuum: general --- techniques: image processing
% Please write all keywords in lower case. PASA uses the
% standard list of subject headings adopted by The Astrophysical Journal
% and available from http://www.journals.uchicago.edu/ApJ/keywords_text.html.
% Keywords are separated by em-dashes, i.e. ---

%%%%%%%%DO NOT EDIT%%%%%%%%%%%%
\medskip
\medskip
\end{minipage}
\end{changemargin}
]
\small
%%%%%%%%EDIT FROM HERE%%%%%%%%%%%%

%Please see the PASA Style Guide for help with correct layout for your manuscript.
%Examples of tables and figures are given below.

\section{Introduction}

The Australian SKA Pathfinder (ASKAP) \citep{Johnston08, Deboer09} is a new radio telescope being built on the Australian candidate Square Kilometre Array (SKA) site in Western Australia. ASKAP will consist of 36 12-metre antennas spread over a region 6 km in diameter. Although the array of antennas is no larger than many existing radio telescopes, each antenna will be equipped with a phased-array feed of 96 dual-polarisation pixels, giving it a 30 deg$^2$ field of view and a very fast survey speed.
The Evolutionary Map of the Universe (EMU) project \citep{Norris11} is a wide-field radio continuum survey planned for ASKAP. The primary goal of EMU is to make a deep
(rms $\sim$ 10$\mu$Jy/bm) radio continuum survey of the entire Southern Sky at 1.3 GHz, extending as far North as $+30^{\circ}$
declination, with a 10 arcsec resolution. EMU is expected to detect and catalogue about 70 million galaxies, including typical star-forming galaxies up to z=1, powerful starbursts to even greater redshifts, and AGNs to the edge of the visible Universe. The amount of data involved with ASKAP ($\sim$2.5 GB/s, or 100 PB/year) requires that the source detection and measurement is fast, robust and highly automated. 

Source detection and measurement is a problem common to all astronomical imaging
surveys and projects, and numerous software tools have been developed in order
perform this initial step in the analysis of imaging data. With the advent of large area
surveys the automation of this process is clearly crucial. This has led to a variety
of survey-specific source-finders being developed, each optimised to address the
specific issues associated with the imaging technology of each survey, and its corresponding
image properties and artifacts. What has not developed in parallel is an analysis of the common
steps in source-identification and measurement in order to assess the optimal
approaches or algorithms that should be used to maximise the robustness and scientific utility of
the resulting source catalogues. This is partially a consequence of images obtained
using different telescope or imaging technologies (UV/optical/NIR/FIR, compared to radio,
or X-ray, or gamma-ray) having very different characteristics. Consequently, the assumptions
being used in source-finders developed for images at one wavelength or technology
are not typically applicable for others.

The analysis presented here looks at the first steps taken in the source-finding process,
those of background estimation and thresholding. This is done explicitly in the context of
radio interferometric imaging, although the expectation is that the conclusions should
be more broadly applicable. Furthermore, our investigation has a focus on identifying an optimal approach for
the source-identification and measurement to be implemented in the ASKAP image
analysis software pipeline. This analysis complements that of Hancock et al.\ 2011 (in prep.),
which presents a detailed exploration of existing source-finding tools, their
underlying algorithms, and how they perform on simulated radio interferometer images.
Together, these analyses comprise the first stage of a thorough investigation of each stage
of the source-identification and
measurement process that is being pursued as part of the Design Study for EMU.

Several radio source identification and measurement tools are in common use.
These include the {\sc Miriad}/AIPS Gaussian fitting routines IMSAD, SAD and VSAD,
{\sc sfind} \citep{hopkins2002}, and Duchamp (Whiting 2008) as well as SExtractor
\citep{bertin1996}. There are also a variety of survey-specific tools, such as HAPPY (a
modified version of SAD used in the FIRST survey, \citealp{white1997}), a machine-learning
back-end to VSAD used to construct the SUMSS catalogue \citep{mauch2003}, BDSM
(used for the LOFAR source-finding, N. Mohan, in prep), and the floodfill algorithm being
used in the Australia Telescope Large Area Survey (ATLAS) Data Release 2
(\citealp{murphy2007}; Hales et al.\ 2011 in prep.).

As standalone tools, none of these is adequate for EMU, due to limitations evident
in the treatment of background estimation, approaches to treating a varying noise level
across an image, and importantly in the numbers of artifacts incorrectly identified as
sources. Here we aim to test the approaches to both background estimation and thresholding,
%with a slight emphasis on the latter,
in order to identify an optimal approach for the EMU survey.
All existing tools have implemented the complete sequence of steps, from background and
noise level estimation, through threshold setting, and ``source-pixel" identification, to
source-measurement. It is thus challenging to extract robust information about each
independent step in the source-identification process. We do this here for a selection
of tools through a judicious choice of parameters in the tasks we investigate, and interpret
our results cautiously as a consequence. We emphasise that we are focussing here on
two-dimensional data (radio continuum images), with our results expected to be applicable
generically to two-dimensional image source-identification.

We present details of the simulated data used in our analysis in \S\,\ref{simdata}, and the
algorithms being tested in \S\,\ref{algorithms}. The background estimation and local rms
noise estimates are discussed in \S\,\ref{bkg}, with the reliability and completeness
statistics being used as a metric to compare the different approaches in \S\,\ref{relcom}.
Our results are summarised in \S\,\ref{conc}.

\section{Simulated Data}
\label{simdata}
\subsection{ASKAP Simulation}

\begin{figure*}[htb] %  figure placement: here, top, bottom, or page
   \centering
   \includegraphics[angle=270,width=3.1in]{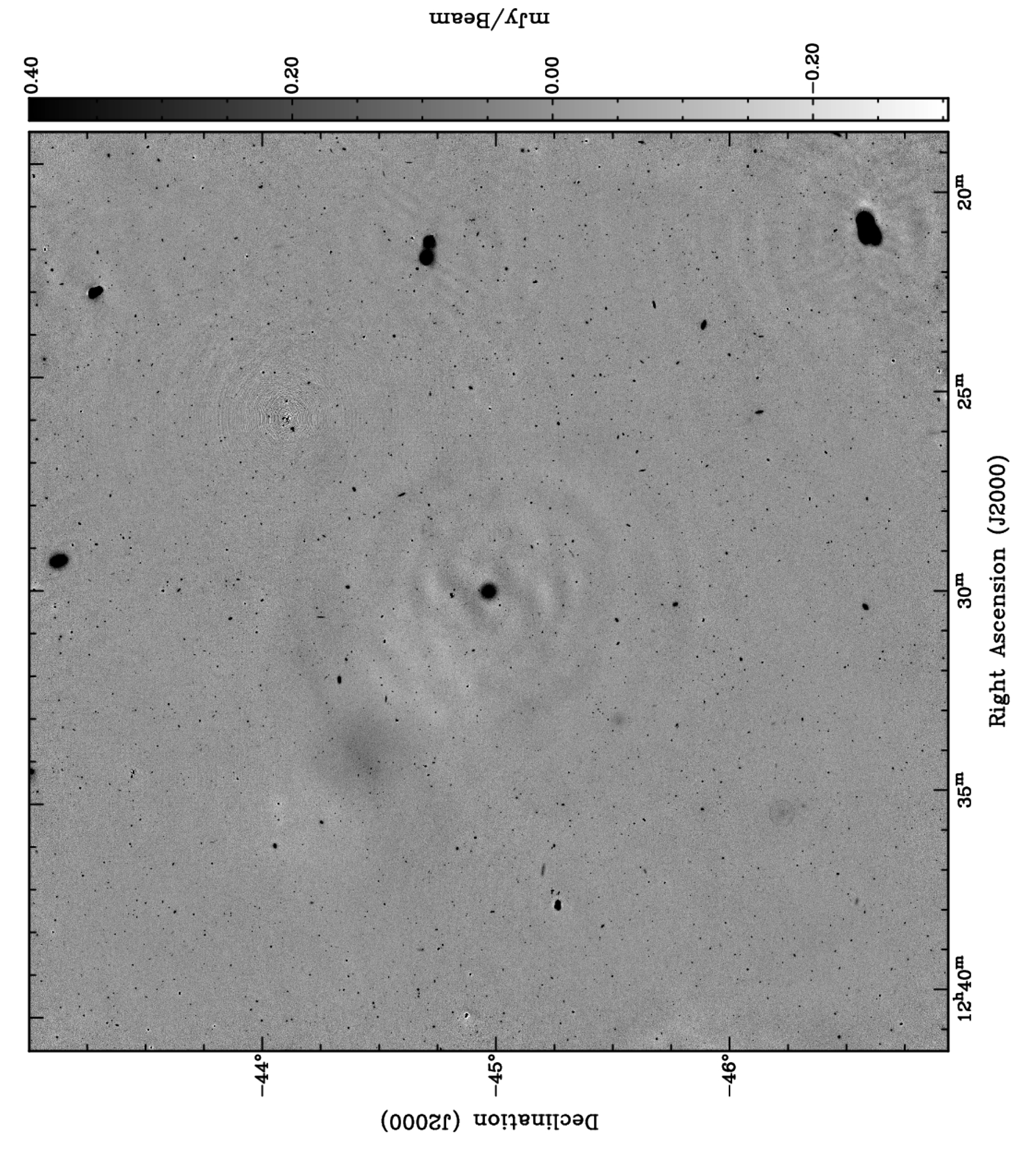} \hspace{3mm}
   \includegraphics[angle=270,width=2.95in]{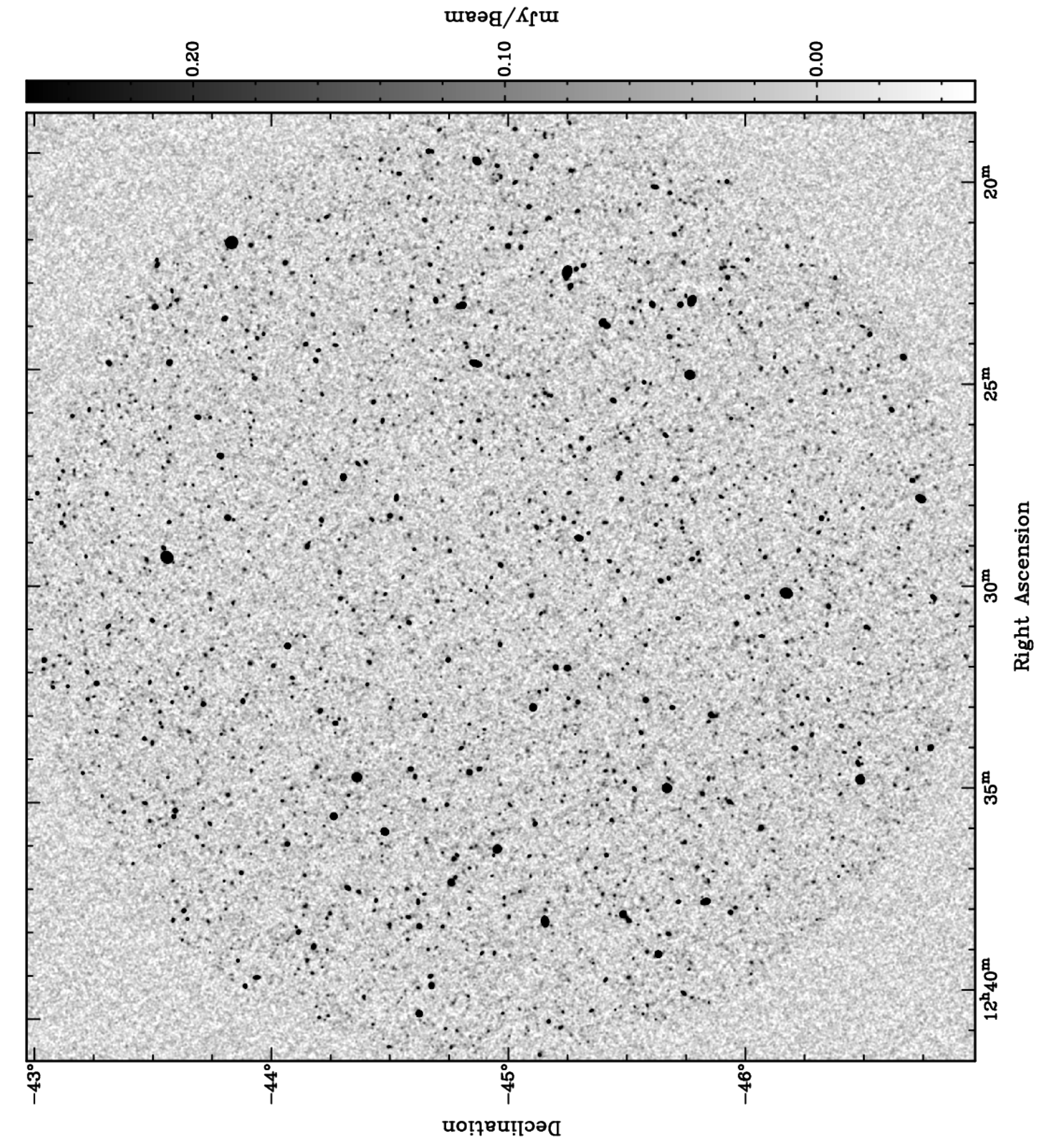} 
   \caption{Left: the ASKAP simulated image with input sources from SKADS S3-SEX source list. Right: the Hancock et al.\ simulated image. Black is positive in this greyscale. Both simulated images are approximately 4 $\times$ 4 degrees in size.}
   \label{fig:sims}
\end{figure*}

We used the December 2010 set of simulations by the ASKAP team, hereafter referred to as the ASKAP simulation,  to test the source extraction methods. This simulation is of a full continuum observation with critically sampled beams and full 6km ASKAP configuration, using an input catalogue of $\sim$ 7.7 million sources down to 1 $\mu$ Jy from the SKADS S3-SEX simulation \citep{wilman2008,wilman2010}.  The effect of the phased-array feed was simulated by using 16 idealised beams, spaced in a rectangular grid one degree apart. This results in 30 -- 40\% peak to peak sensitivity variation in the simulation. The pixel scale of this simulation is 2.75 arcsec. For these tests we removed the outer 500 pixels of the simulated continuum to reduce potential edge effects.
%or aliasing present in the simulation.
%For example, the previous version of the simulations (SST3), had significant aliasing effects which resulted in bright sources in the image even though they weren't in the input source list. 
The simulated continuum image used in these tests is shown in Figure \ref{fig:sims}. The noise in the simulated image is approximately 35 $\mu$Jy rms, although it varies across the field.  

\subsection{Hancock et al.\ Simulation}

The ASKAP simulation is made up of millions of sources and includes instrumental artefacts. The SKADS S3-SEX input source list contains many multiple sources (FRI and FRII galaxies) as well as extended sources. We wish to test the source extraction algorithms where deblending and extended sources are not an issue. We also wish to test the algorithms for an idealised case of Gaussian noise. For this we use the simulation of Hancock et al.\ 2011 (in prep). In this simulation a catalogue of sources was created with the flux of the sources drawn from the source count distribution $N(S) \sim S^{-2.3}$ \citep{hopkins1998}. The sources are all compact (1 -- 2 FWHM) when convolved with the synthesised beam (30 arcsec). A map with 25 $\mu$Jy Gaussian noise was then created and convolved with a 30 arcsec beam, before the sources were injected into the image. The image has a pixel scale of 6 arcsec so that the synthesised beam of the telescope is sampled 5 times in each direction. There are 15,000 sources in the simulated image with fluxes $>1\,\sigma$. 

\section{Source Extraction Algorithms}
\label{algorithms}

A review of existing source-finding tools, their underlying algorithms, and how they perform on simulated radio interferometer images, is presented by Hancock et al.\ 2011 (in prep).  Hancock et al., (2011, in prep) compares the performance
of the source-finders {\sc imsad}, {\sc sfind}, SExtractor, Selavy, Floodfill and a newly developed tool, Tesla, in the identification and characterisation of artificial sources. Their analysis emphasises some common failure modes in existing tools, and identifies a potential solution, implemented and testing using the new tool, Tesla, to rectify these shortcomings.

To complement the analysis of Hancock et al. (2011, in prep), we focus here on the background estimation
background estimation and thresholding approaches as implemented
in the SExtractor, Duchamp and {\sc sfind} algorithms for source extraction. SExtractor and Duchamp
implement a fixed signal-to-noise (S/N) thresholding technique while {\sc sfind} uses a
statistical method called false-discovery rate (FDR) that sets a threshold based on a user-specified
limit to the fraction of falsely detected sources \citep{miller2001,hopkins2002}. The thresholds set by these algorithms are based on peak flux densities, not integrated flux densities.

SExtractor \citep{bertin1996} is a source extraction tool that detects sources through
thresholding. A group of connected pixels brighter than some threshold above the background is identified as a detection. SExtractor uses several steps to detect sources. These are background subtraction, image filtering, thresholding, deblending, and source parameterization (including isophotal analysis, photometry and astrometry). 

Duchamp (Whiting 2011, in prep) is a source-finding tool designed for use with spectral-line cubes, particularly those dominated by noise with relatively small sources present (for example extragalactic HI surveys or maser surveys). Sources are identified by applying a threshold (a uniform one for the entire image/cube) and grouping sets of adjacent (or suitably close) voxels together. It is possible to do various types of pre-processing to enhance the detectability of sources (e.g. smoothing, spectral baseline subtraction, wavelet reconstruction).  Duchamp forms the basis for the prototype source finder for the ASKAP science processing pipeline. This implementation, known as Selavy to distinguish it from the standalone Duchamp, is still under development. It has several features that do not appear in Duchamp. These are described in the paper by Whiting (2011) in this volume. The key feature of these, relevant to this work,
is the ability to vary the threshold according to the local noise properties. This uses a similar procedure to SExtractor: defining a set box-size, and, for each pixel, finding the noise properties within a box centred on that pixel. In this way, a different flux threshold can be defined for each pixel, given a signal-to-noise ratio threshold. This work makes use of the Selavy implementation, to take advantage of this and other new features, and we refer to this algorithm as Selavy hereafter.  

The false-discovery rate (FDR) is a statistical procedure which is an alternative to the simple threshold
definition used in the identification of sources. In SExtractor or Selavy (and many traditional
approaches to source-identification) a source is initially identified by pixels with an intensity
above a threshold defined as some multiple of the local rms noise level, x$\sigma$. In the
FDR approach, the threshold is defined through a robust statistical procedure that takes into
account the intensity distribution of all pixels in an image (both source and noise), compared
to an image of equal size containing only noise, in setting a threshold. The resulting threshold
places a limit on the fraction of sources identified that may be false based simply on the
statistics of the noise distribution \citep{miller2001}. This procedure has been implemented
for source measurement in radio images in the M{\sc iriad} task {\sc sfind} (Hopkins et al.\ 2002) but while it is also an option for defining the
threshold in the stand-alone version of Duchamp, it is not
yet implemented in Selavy.
The key parameters for {\sc sfind} in implementing the FDR algorithm are the rmsbox size (similar to the mesh-size for Sextractor) and $\alpha$, which is the desired fractional limit
to the number of false detections in the final source list.

\section{Background and Noise Maps}
\label{bkg}

The first stage of source extraction in general is background estimation. In most astronomical images the background is non-zero and varies over the frame. 
Radio images from interferometeric synthesis techniques may in general have a non-zero background, although in many cases it is small (e.g. \citealp{rich2008}).
It is not expected that the EMU images will have a significant non-zero background, but the background estimation step is nevertheless important.
In the first place, a significant non-zero background may be an indicator of problems with
the data or the observation, and can be used as a step in quality control of the imaging data.
Moreover, there will be low level diffuse emission close to the galactic plane, and extended structures such as supernova remnants at latitudes up to tens of degrees. For the identification and
measurement of point-sources, these extended diffuse structures and emission can be treated
as a background and removed in the same fashion. The Planck Early Release Compact Source Catalogue (Planck Collaboration 2011) demonstrated that extragalactic radio sources can still be extracted successfully in the galactic plane with careful background estimation. 
Regions away from the plane of the Milky Way are expected to have a zero background

It is also a common technique to apply some filtering or weighting of images before
the background and noise estimation is performed. SExtractor allows both filtering and weighting,
and Selavy allows filtering, but {\sc sfind} does not perform either filtering or weighting. 
These aspects of background estimation are not explored in
the current analysis, in order to enable more direct comparison of the actual background
estimation algorithms being applied.

\subsection{SExtractor}
The value measured at each pixel of a radio image is a sum of the background signal and
the emission from the radio sources of interest. To construct a background map SExtractor computes an estimator for the local background in a rectangular region, or ``mesh", of a grid that covers the whole image. The estimator is a combination of $\kappa \sigma$ clipping and mode estimation (similar to DAOPHOT, see e.g. \citealp{dacosta1992}). In brief, the local background is clipped iteratively until convergence at $\pm3\sigma$ around its median and then the mode is estimated. The background map is generated from a bicubic-spline interpolation between the meshes of the grid. 

The choice of mesh-size is very important. If it is too large then small scale variations in the background and noise will be lost, but if it is too small then the background and noise estimations will be affected by object emission. Published surveys have found mesh-sizes with widths of 8 to 12 times the
point-spread-function (PSF), or synthesised beam size, produced good results for noise estimation in
deep radio continuum surveys \citep{huynh2005,schinnerer2007,schinnerer2010}.  In the Galactic plane, the Planck Early Release Compact Source Catalogue (Planck Collaboration 2011) found empirically that a background mesh-size of 4 to 24 beamwidths yields a background image that successfully combines substructure in the background and the instrumental noise in the image.

In this study we explore mesh-sizes of 3, 10, 20 and 100 times the beamwidth. SExtractor also has a smoothing parameter which allows the background map to be smoothed to suppress any local over-estimations due to bright sources, hence a mesh-size of 10 beamwidths with a smoothing filter of 3 mesh-sizes (30 beamwidths) is also investigated. The difference between choosing a large mesh-size and smoothing on the same scale is subtle. Choosing a mesh-size of 30 beamwidths splits the image into a grid of 30 $\times$ 30 beamwidth squares, the background level and rms is calculated in those regions, and then the background and noise images are made from a bi-cubic spline fit to this grid of statistics. Smoothing, on the other hand, is a median filter (here 30 beamwidths) applied after the background and noise image is made from the smaller mesh-size (here 10 beamwidths). Smoothing is therefore useful where the bi-cubic spline interpolation breaks down, such as crowded fields.
Initial investigation showed that mesh-sizes of 10 and 20 beamwidths produce satisfactory background and noise images, and subsequent investigation of reliability and completeness will focus on those, as well as that with the smoothing filter applied.

\begin{figure*}[htb] %  figure placement: here, top, bottom, or page
   \centering
    \includegraphics[width=5.8in]{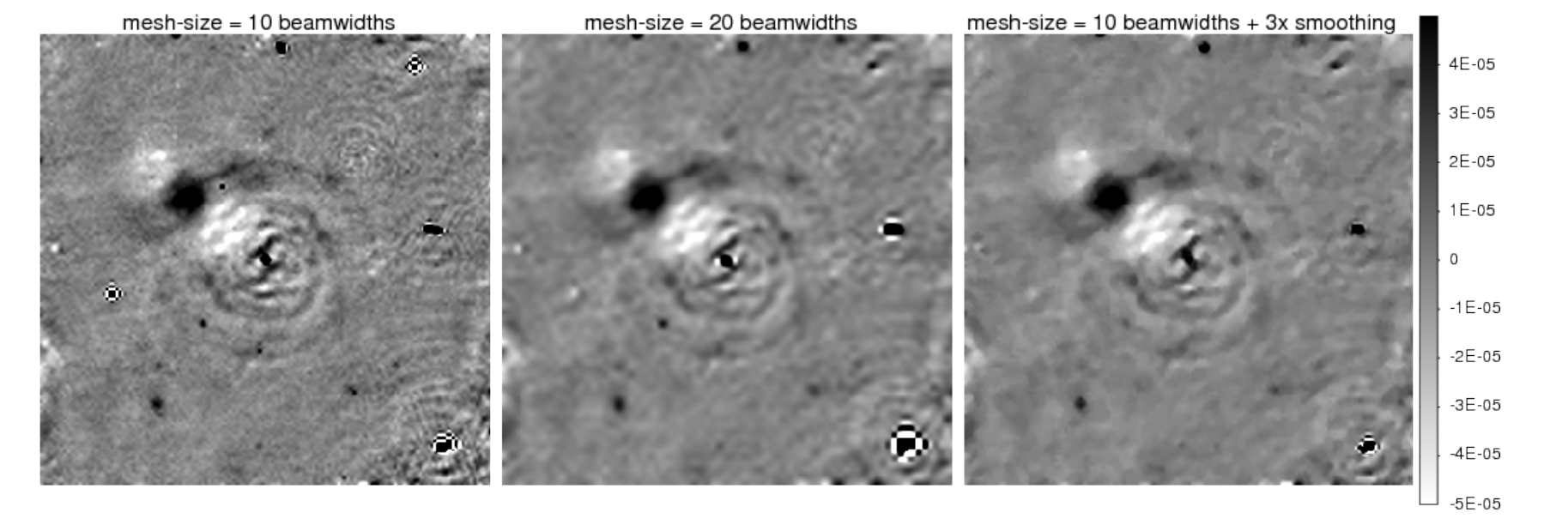} 
   \caption{The SExtractor background estimates of the ASKAP simulated image. The left and center images have mesh-sizes of 10 and 20 beamwidths, respectively, and no smoothing. The right image has a mesh-size of 10 beamwidths and a smoothing scale of 3 mesh elements.
   The greyscale is $-50$ to $50\,\mu$Jy in all images. Black is positive in this greyscale.}
   \label{fig:bg}
\end{figure*}

\begin{figure*}[htb] %  figure placement: here, top, bottom, or page
   \centering
   \includegraphics[width=5.8in]{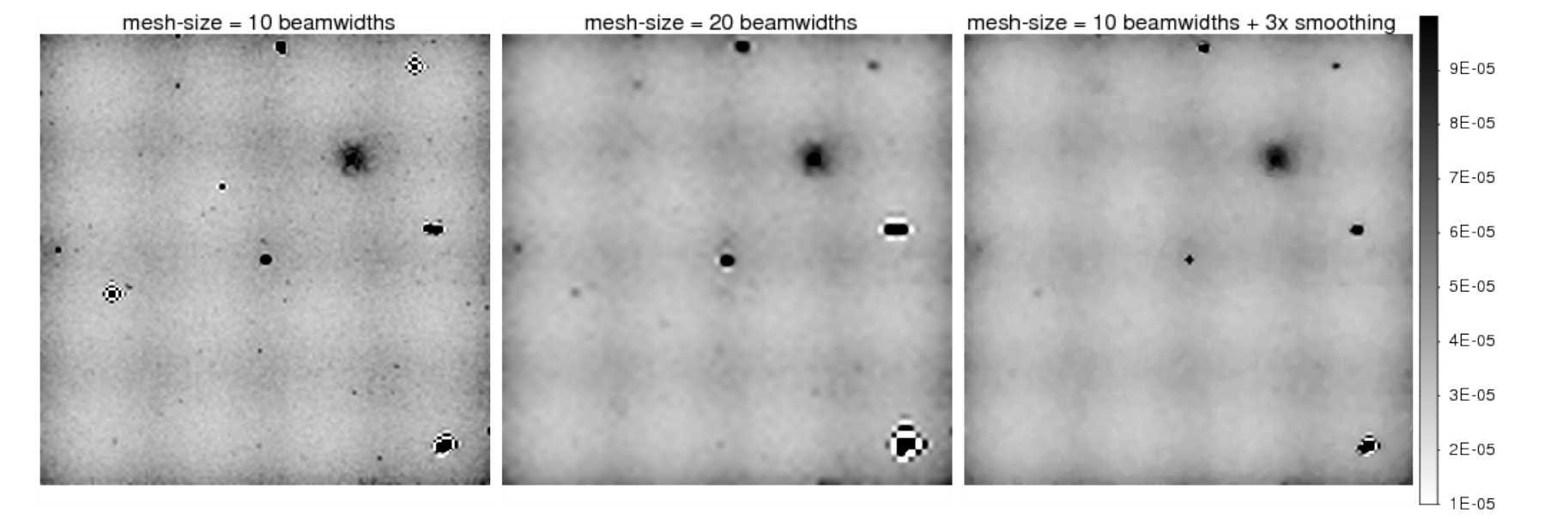}
   \caption{The SExtractor noise image calculated from the ASKAP simulated image. The left and center images have mesh-sizes of 10 and 20 beamwidths, respectively. The right image has a mesh-size of 10 beamwidths and a smoothing scale of 3 mesh elements. The greyscale is $10$ to $100\,\mu$Jy in all images. Black is positive in this greyscale.}
   \label{fig:rms}
\end{figure*}

\begin{figure*}[htb] %  figure placement: here, top, bottom, or page
   \centering
    \includegraphics[width=5.8in]{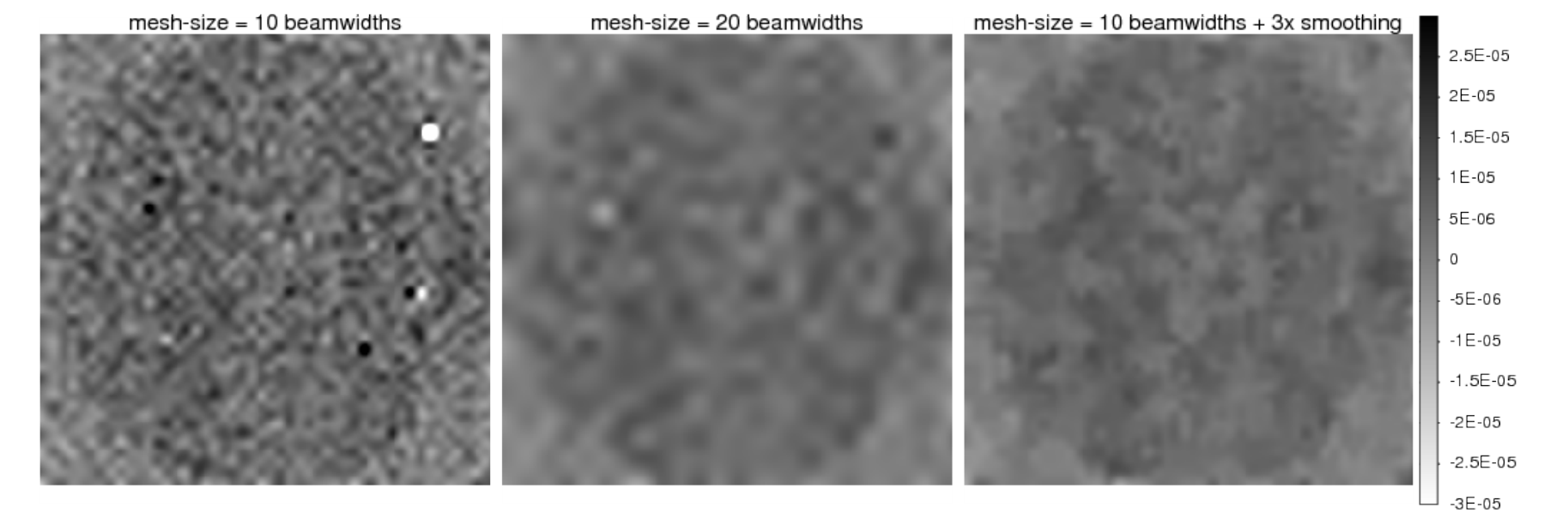}
       \caption{The SExtractor background estimates of the Hancock et al.\ simulated image. The left and center images have mesh-sizes of 10 and 20 beamwidths, respectively, and no smoothing. The right image has a mesh-size of 10 beamwidths and a smoothing scale of 3 mesh elements.
       The greyscale is $-30$ to $30\,\mu$Jy in all images. Black is positive in this greyscale.}
   \label{fig:bg2}
\end{figure*}

\begin{figure*}[htb] %  figure placement: here, top, bottom, or page
   \centering
   \includegraphics[width=5.8in]{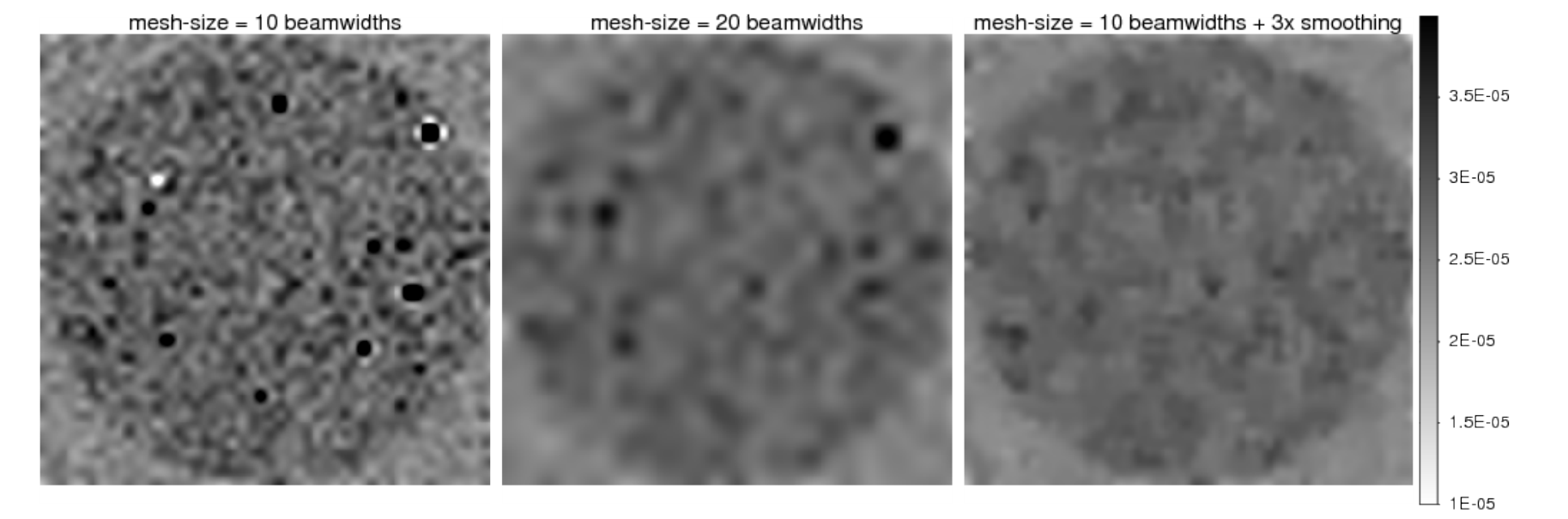}
      \caption{The SExtractor noise image calculated from the Hancock et al. simulated image. The left and center images have mesh-sizes of 10 and 20 beamwidths, respectively. The right image has a mesh-size of 10 beamwidths and a smoothing scale of 3 mesh elements. The greyscale is $10$ to $40\,\mu$Jy in all images. Black is positive in this greyscale.}
   \label{fig:rms2}
\end{figure*}

The background and rms noise images for mesh-sizes of 10, 20 and 30 (10 smoothed by 3) beamwidths are shown in Figures \ref{fig:bg} to \ref{fig:rms2}. All background maps show large scale ripples in the image from sidelobes near the brightest sources in the ASKAP simulations. The rms noise maps show high noise regions around these brightest sources, as desired for spurious source rejection. The rms noise map, however, is affected by fainter ($\sim$ 1 to 10 mJy) point sources when the mesh-size is only 10 beamwidths (Figures \ref{fig:rms} and \ref{fig:rms2}). These sources are well-cleaned and do not have significant sidelobes, so they should not be contributing to the noise in the image. We therefore find that using a mesh-size of 10 beamwidths but smoothing to 30 beamwidths seems to result in the highest quality background and rms maps. The effect of the phased array feed on the ASKAP simulations can be seen in the grid pattern of low and high noise regions in the background images (Figure~\ref{fig:rms}). 

\subsection{SFIND}

{\sc sfind} determines the fraction of expected false sources by first estimating the background and
rms for the whole image using uniformly distributed regions of a user-specified ``rmsbox" size in pixels.
A normalised image is made from the input image by subtracting the mean and then dividing by the rms. If the image noise properties are Gaussian, then the normalised image would ideally
show a Gaussian distribution with a mean of 0 and $\sigma=1$. The false-discovery rate method is implemented on this normalised image. Each pixel is assigned a $p$-value, a probability that it is
drawn from the noise distribution, from this image. Therefore the quality of this normalised image is important and the ``rmsbox" parameter needs to be carefully considered.

\begin{figure*}[htb] %  figure placement: here, top, bottom, or page
   \centering
   \includegraphics[width=2.55in]{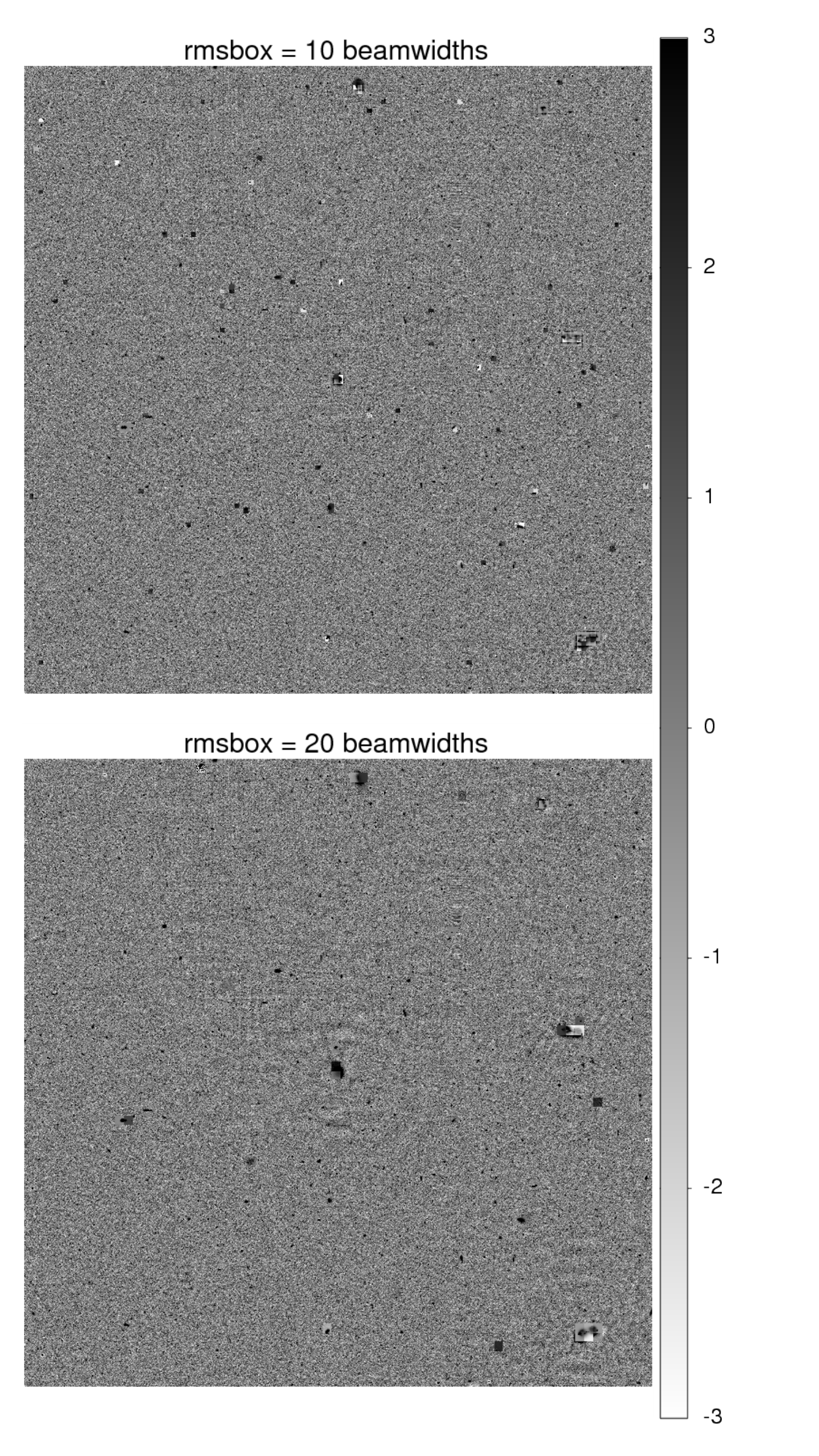}   \includegraphics[width=2.5in]{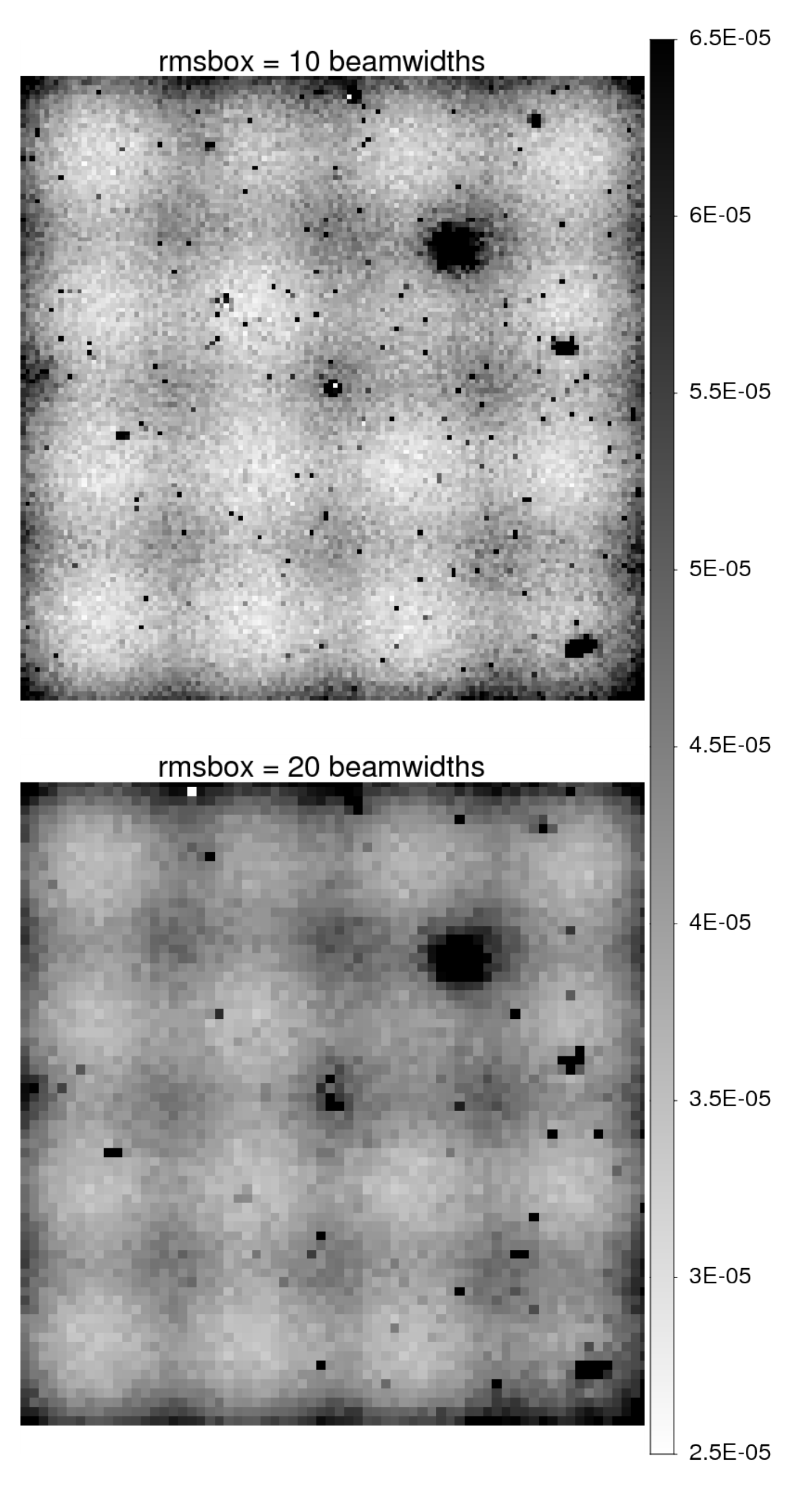}
   \caption{The {\sc sfind} normalised (left) and rms (right) images calculated from the ASKAP simulated image. Top row is for a 10 beamwidth rmsbox, and the bottom is for a 20 beamwidth rmsbox. The greyscale of the normalised images is $-3$ to $3\,\sigma$. The greyscale of the rms images is $25$ to $65\,\mu$Jy. Black is positive in this greyscale.}
   \label{fig:sfindbg}
\end{figure*}

\begin{figure*}[htb] %  figure placement: here, top, bottom, or page
   \centering
   \includegraphics[width=2.5in]{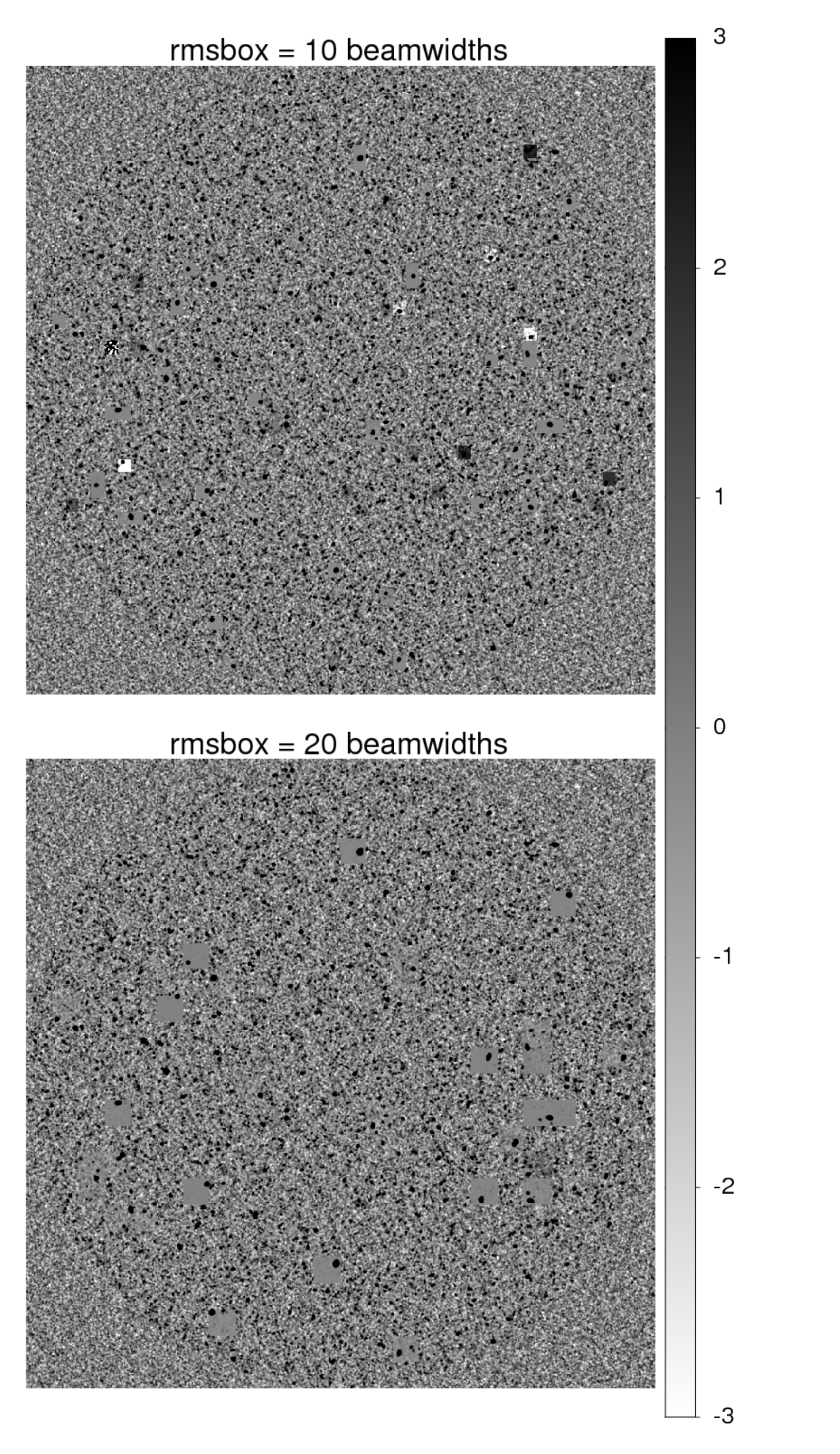}\includegraphics[width=2.5in]{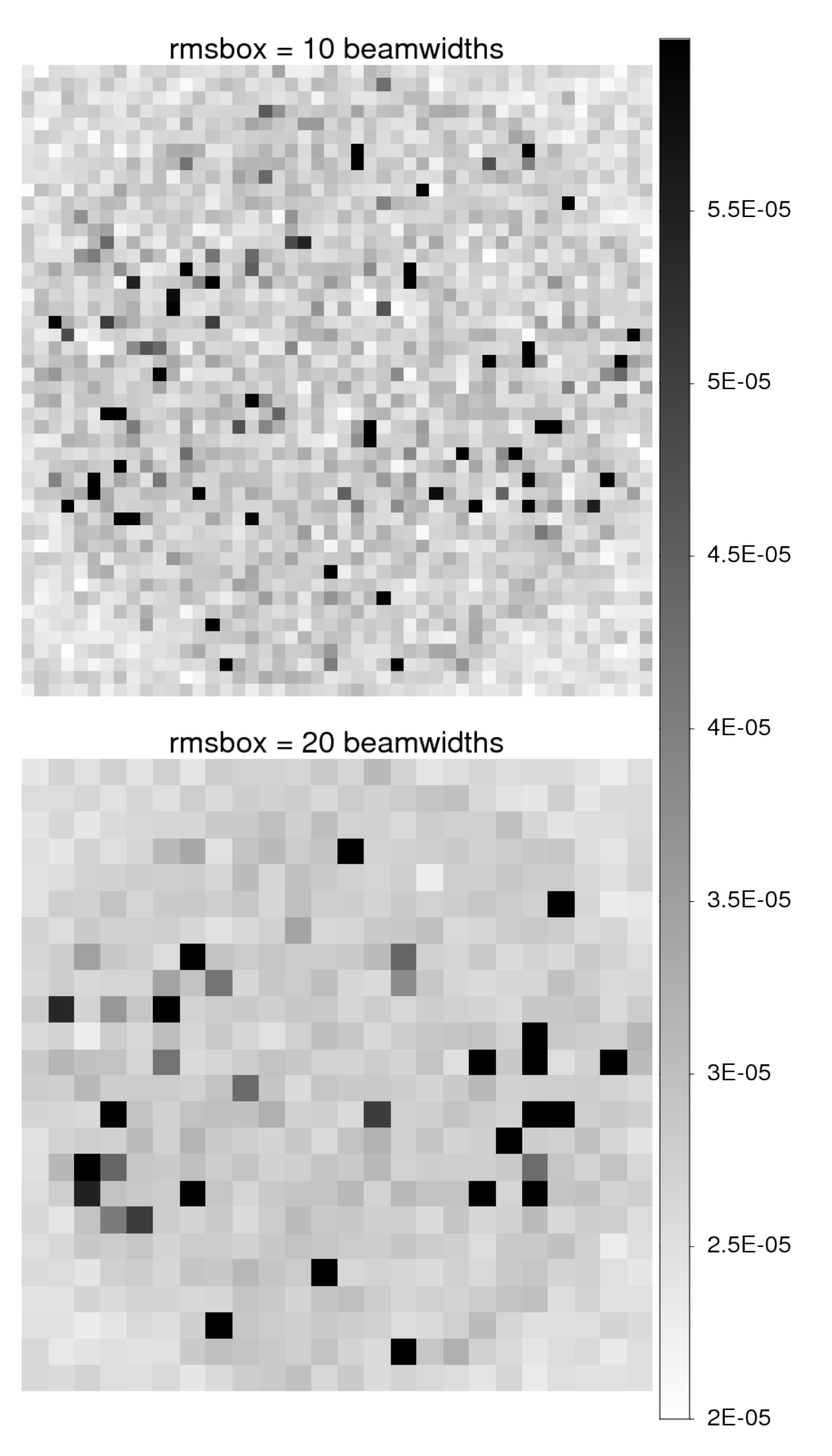}
   \caption{The {\sc sfind} normalised (left) and rms (right) images calculated from the Hancock et al.  simulated image. Top row is for a 10 beamwidth rmsbox, and the bottom is for a 20 beamwidth rmsbox. The greyscale of the normalised images is $-3$ to $3\,\sigma$. The greyscale of the rms images is $20$ to $60\,\mu$Jy. Black is positive in this greyscale.}
   \label{fig:sfindbg2}
\end{figure*}

The output normalised and rms images for rmsbox sizes corresponding to 10 and 20 beamwidths are shown in
Figures~\ref{fig:sfindbg} and \ref{fig:sfindbg2}. For both rmsbox sizes, the ASKAP normalised images have a mean of 0.03 and standard deviation of 1.04, when pixels with an absolute value greater than 5 are excluded from the statistics. Similarly, for the Hancock et al. simulations a rmsbox size of 10 beamwidths results in a mean of 0.10 and standard deviation of 1.10, while 20 beamwidths results in a mean of 0.10 and standard deviation of 1.08. So the normalised images show close to the ideal Gaussian distribution of 0 mean and $\sigma=1$. 

The rms image from a rmsbox size of 10 beamwidths appears to be more affected by bright sources than rms image using the larger rmsbox of 20 beamwidths. 
As with the SExtractor rms maps, high noise estimates are expected around the brightest sources with significant sidelobes, and this is necessary to accurately reject false detections.
Sources as faint as a few mJy affect the noise image for a rmsbox size of 10 beamwidths (Figures~\ref{fig:sfindbg} and \ref{fig:sfindbg2}), but this is not desirable as these sources do not have significant sidelobes. 

\subsection{Selavy}

As with the other source extraction algorithms, Selavy requires the background and noise levels to determine a threshold level. It does this by calculating the mean or median for the background, and rms or the median absolute deviation from the median (MADFM) for the noise level. The median and MADFM are robust statistics which are not biased by the presence of only a few bright pixels, and are used by Selavy by default. The ASKAP software pipeline implementation of Selavy used in this work does not yet allow the output of the background and noise images, so no direct comparison of these images can be made. 

\subsection{Results of background and noise estimation}

\begin{figure*}[htb] %  figure placement: here, top, bottom, or page
   \centering
   \includegraphics[width=3.1in]{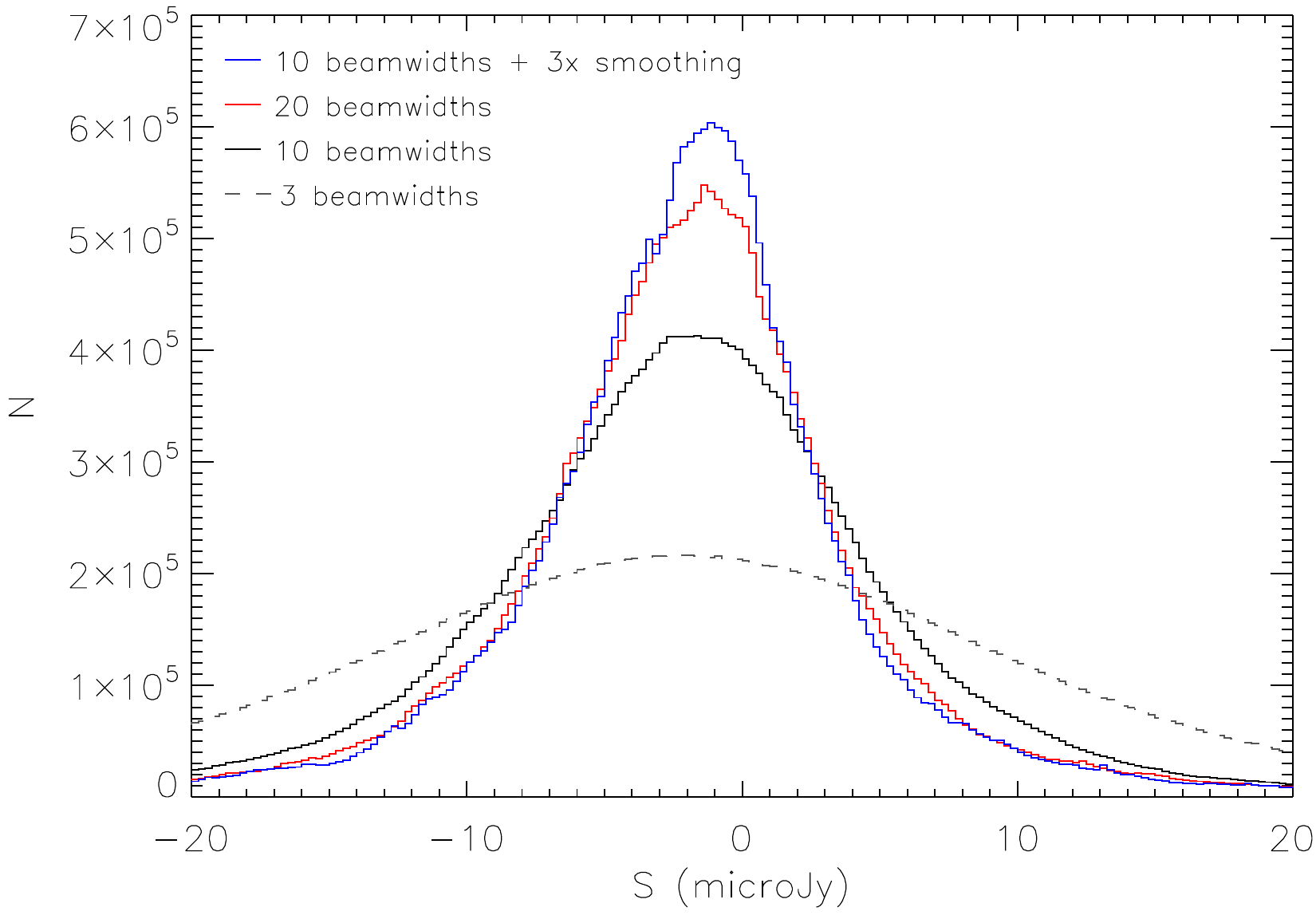} 
    \includegraphics[width=3.2in]{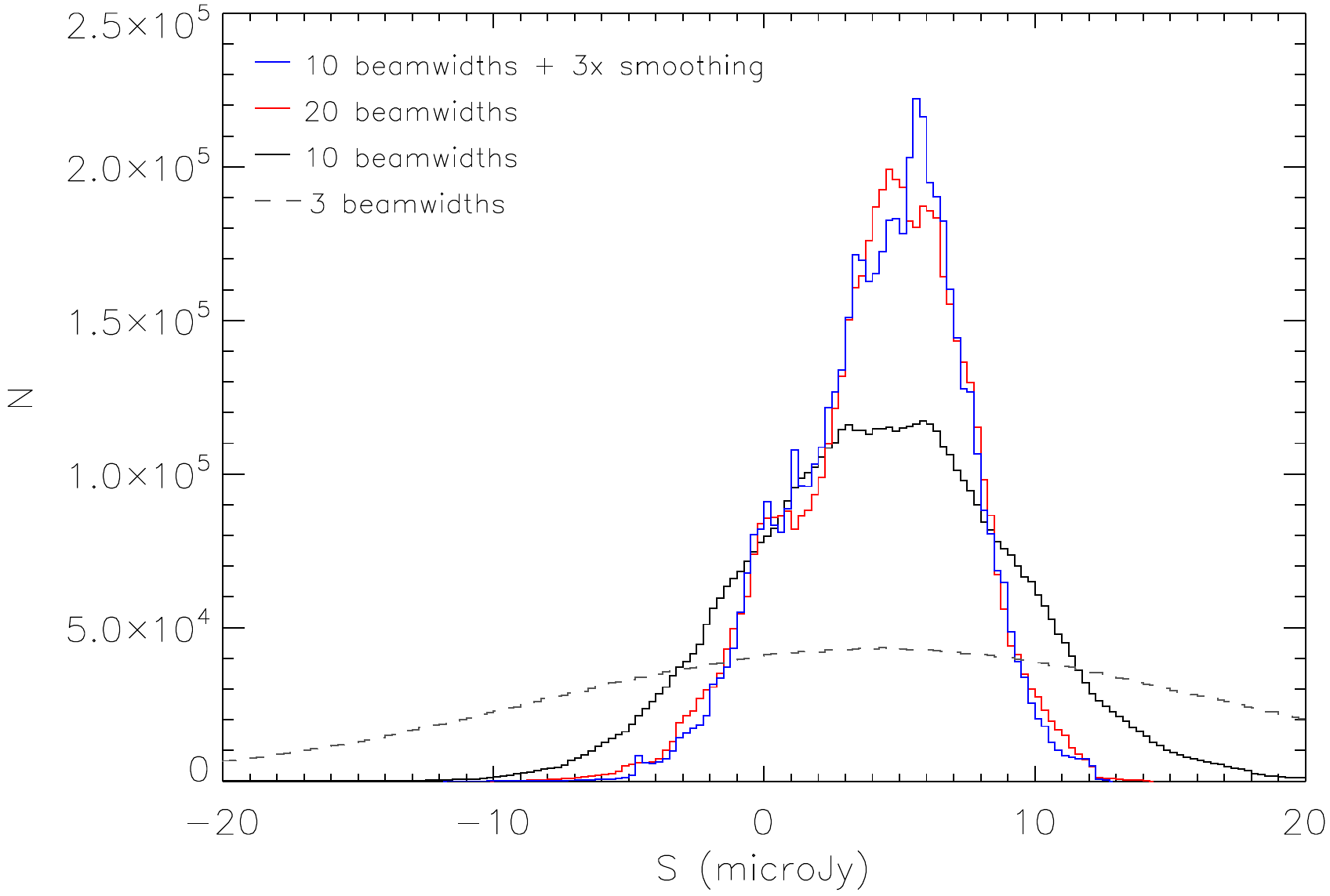} 
   \caption{Histogram of the pixel values in the ASKAP background image (left) and Hancock et al.\
   background image (right) generated by SExtractor. Black and red lines are for 10 and 20 beamwidth mesh-sizes, respectively. The blue line is for a mesh-size of 10 beamwidths and a smoothing scale of 3 mesh elements. Grey dashed line is for a mesh-size of 3 beamwidths.}
   \label{fig:bgplot}
\end{figure*}

To examine the background and noise images quantitatively we examined the distribution of pixel values of these images. A mesh-size of 10 beamwidths results in a larger spread in background values than either 20 beamwidths or 10 beamwidths plus smoothing (Figure \ref{fig:bgplot}). All mesh-sizes give a similar median pixel value in the background image, i.e.\ similar average background across the whole image. For comparison we also show the pixel distribution for a mesh-size of 3 beamwidths (grey dotted line in Figure \ref{fig:bgplot}). This also has a similar median pixel value, but a much greater variation in background levels, indicating that the background determination is probably affected by local structure (e.g. radio sources, local noise peaks and troughs) for this small mesh-size. 

\begin{figure*}[htb] %  figure placement: here, top, bottom, or page
   \centering
   \includegraphics[width=3.1in]{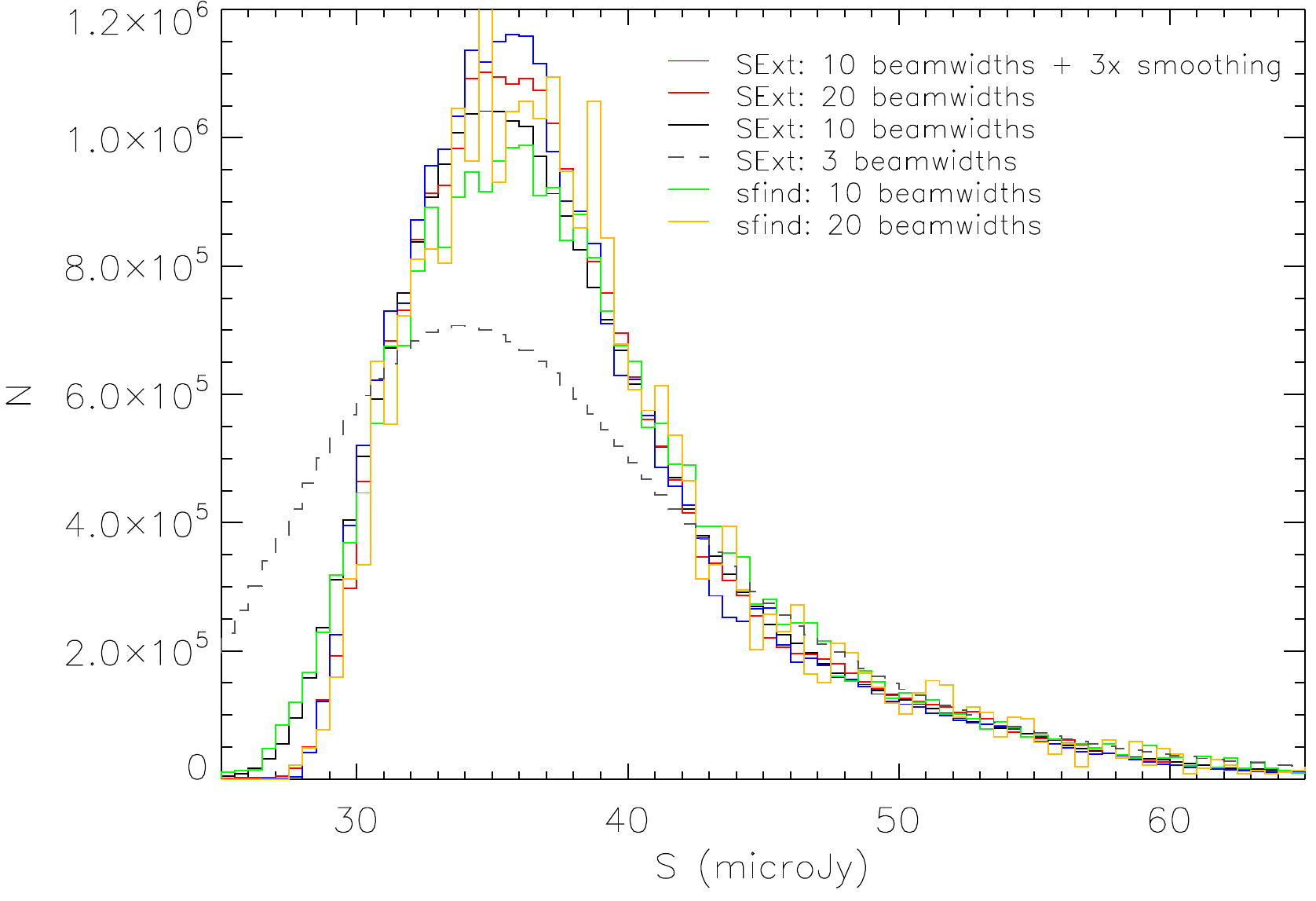} 
    \includegraphics[width=3.07in]{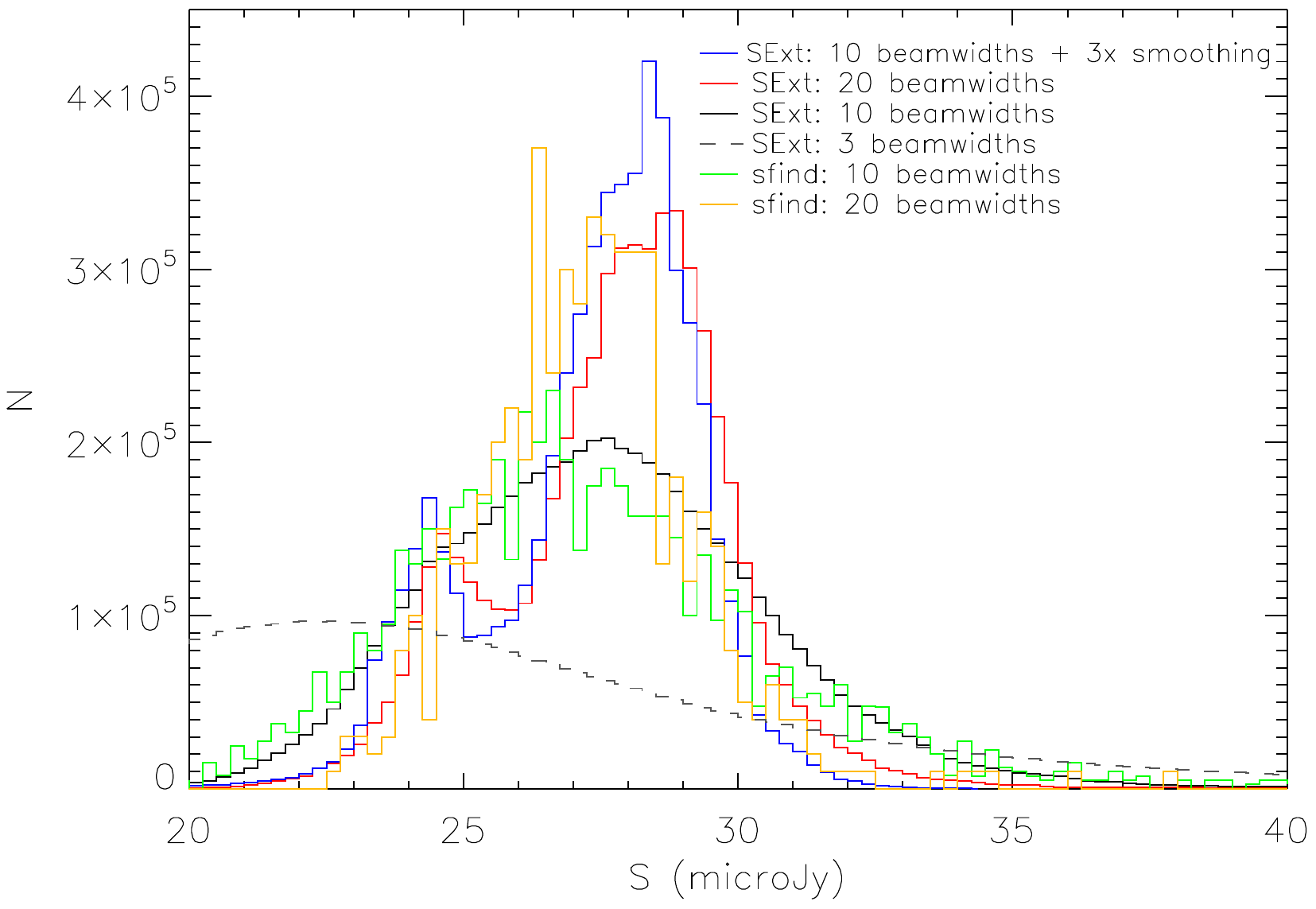} 
   \caption{Histogram of the pixel values in the ASKAP rms images (left) and Hancock et al. rms images (right) generated by SExtractor and {\sc sfind}. Black and red lines are for SExtractor with 10 and 20 beamwidth mesh-sizes, respectively. The blue line is for SExtractor with a mesh-size of 10 beamwidths and a smoothing scale of 3 mesh elements. The grey dashed line is for SExtractor with a mesh-size of 3 beamwidths. The green and gold lines are for {\sc sfind} with rmsbox of 10 and 20 beamwidths, respectively.}
   \label{fig:rmsplot}
\end{figure*}

The pixel distribution of the rms noise images constructed by SExtractor and {\sc sfind} are shown in Figure \ref{fig:rmsplot}. 
In the case of the ASKAP simulations there is no significant difference between using a 10 beamwidth mesh-size and 20 beamwidth mesh-size, with only minor differences appearing
in the distributions seen at the lowest rms noise levels, below about $30\,\mu$Jy. The rms
noise level of the ASKAP simulation is approximately $35\,\mu$Jy, although with 30--40\%
variations over the field. Both SExtractor and {\sc sfind} appear to be recovering
noise estimates that peak around $35\,\mu$Jy, for both mesh-sizes tested. 

In the case of the Hancock et al.\ simulation, however, there is a significant difference in noise image
pixel values for mesh-sizes of 10 versus 20 beamwidths. The input rms noise level for this image
was $25\,\mu$Jy, but is effectively a few percent higher due to the convolution with a Gaussian 
beam leading to correlated noise between pixels. All estimates seem to be slightly higher than $25\,\mu$Jy, 
as expected, but with those from {\sc sfind} being marginally (but systematically) lower than those from SExtractor.
We also see a bimodal distribution in the noise image pixel values for some cases (20 beamwidths and smoothing to 30 beamwidths). This is because the convolution with a Gaussian was only applied to a circular region of interest (as seen in Figure 1). The empty (noise-only) areas in the full Hancock et al. image which lie outside this region remain included in these statistics, however, leading to the secondary 'peak' near $25\,\mu$Jy.
The 20 beamwidth mesh-sizes in both cases show more narrowly-peaked distributions than those
of 10 beamwidths, and with the peak value at higher flux densities. This is likely to lead, for the
20 beamwidth mesh-sizes, to fewer spurious detections (a higher reliability) for a given threshold
level, but possibly at the expense of completeness.

The pixel distribution for an SExtractor mesh-size of 3 beamwidths peaks at a substantially lower
value for both simulations, indicating underestimated rms noise values. The reliability estimates in
\S\,\ref{relcom} below imply that such a small mesh-size and the correspondingly lower
rms noise values inferred leads to an increased number of spurious sources being detected.

Overall, the background estimation behaviour by both SExtractor and {\sc sfind}, for common
mesh-sizes in each simulation, was similar. It is clear, though, that the choice
of mesh-size can lead to substantially different background estimates, and the fact that
this is typically a user-defined step in existing source-finding tools suggests that it is perhaps
a challenging parameter to estimate in an automated fashion.

\section{Reliability and Completeness}
\label{relcom}

The efficacy of different approaches to background estimation and threshold setting is
being assessed in our analysis by reference to the completeness and reliability statistics
for the resulting source catalogues, compared to the known input catalogues. Here we
present the details of these metrics for different combinations of the approach to
the background estimation and the threshold level, for each of the algorithms being explored,
as implemented in the three software tools.

Completeness is a measure of the fraction of real sources detected. For our analysis
we define completeness as the fraction of {\em input\/} sources which have a detected
{\em output\/} source counterpart. Input sources were deemed to be detected if they had at least
one measured source within 0.5 synthesised beam FWHM.

Reliability is a measure of the fraction of detected sources that are real.
For our analysis we define reliability as the fraction of {\em output\/} sources which have an
{\em input\/} catalogue counterpart. The input source list was limited to those artificial sources having
a flux density greater than $1\,\sigma$, in order to keep the number of ``truth" sources to a
manageable quantity. Sources fainter than $1\,\sigma$, (which exist in the input SKADS S3-SEX
simulated source list for the ASKAP image), are much fainter than the lowest tested threshold
($3\,\sigma$) for our analysis. We do not expect that the omission of these from our analysis will
have any effect on our results.

The measured sources were deemed to have a
counterpart based only on positional coincidence, specifically if there is at least one ``truth" source
within 0.5 synthesised beam FWHM. We place no constraint on how the output flux density compares
to the ``truth" value, although this is another criterion that has been applied by other
teams in determining whether a source has been recovered successfully (e.g.\ Planck ERCSC team).
Here we are not exploring the properties of the source fitting routines, limiting ourselves only
to the statistics of detections as a metric for assessing the background and noise estimation
algorithms. Consequently we do not include the output flux density estimates in
assessing the recovery of input artificial sources.

\subsection{SExtractor}
\label{subsec:sex}

SExtractor detects, and performs source-measurement on, pixel islands that lie above the user given threshold, which we set as a multiple of the noise image calculated in Section 3.1. This is in effect applying a local S/N threshold, provided the rms noise image is accurate. Our goal is to estimate the reliability and completeness of the source extraction for various thresholds. 

The reliability for SExtractor on the ASKAP simulation, for thresholds of $3\,\sigma$ to $20\,\sigma$,
is summarised in Table~\ref{tab:rel1}. The reliability rises sharply between
$3\,\sigma$ and $5\,\sigma$ from 54\% to 88\%. There is
little difference in results from noise maps made from a 10 beamwidth mesh-size compared to that
made with a 20 beamwidth mesh-size or with a 3 mesh-size smoothing.
The $15\,\sigma$ and $20\,\sigma$ threshold SExtractor catalogues have surprisingly low
reliabilities of only 92\% and to 94\%, and it turns out that this is due to a deblending issue.
Multiple bright sources may be extracted as a single source with a position halfway between
the input sources, hence they have no ``truth" counterpart. This effect is exacerbated when
the threshold level is set so high, and a relatively large fraction of the input sources fall into
this scenario.

The reliability for SExtractor on the Hancock et al.\ simulation is summarised in Table~\ref{tab:rel2}.
The reliability rises sharply between $3\,\sigma$ and $4\,\sigma$, from 88\% to 98\% in the case
of a 10 beamwidth mesh-size. Again, there is no significant difference to the reliability from
applying the larger mesh-size or smoothing in the noise calculation. The reliability is almost
100\% by a threshold of $5\,\sigma$, which is expected in this simulation where the noise is
completely Gaussian and the ``truth" catalogue is known. 

The completeness as a function of input source flux density is shown
in Figure~\ref{fig:comp}, for SExtractor thresholds of 3, 4, 5, and $10\,\sigma$. 
A known bug in {\sc sfind} limits results to an rmsbox size of 10 beamwidths (see \ref{subsec:sfindcomp}), so we show results from a mesh-size of 10 beamwidths to allow a comparison across all 3 algorithms.
The completeness increases significantly at the faint flux density levels going from $10\,\sigma$ to $3\,\sigma$, as expected. The completeness does not reach unity, however, even for low thresholds of $3\,\sigma$
and $4\,\sigma$, for the ASKAP simulation. Since the noise is $\sim 40\,\mu$Jy in the ASKAP simulation we expected completeness to be about 100\% by $\sim 400\,\mu$Jy for these detection thresholds. In comparison, SExtractor performs as expected with the Hancock et al. simulations and in this
``perfect" image the completeness reaches 100\% for $3\,\sigma$ and $4\,\sigma$ thresholds at
about $300\,\mu$Jy. 

Tables~\ref{tab:comp} and \ref{tab:comp2} summarise the completeness for various flux density bins and SExtractor mesh-sizes for a threshold of $3\, \sigma$. For both simulations the completeness increases in the lowest flux density bins as the mesh-size is decreased to 3 beamwidths. This is consistent with the lower reliability of sources for this small mesh-size. In this case more sources are extracted, hence the greater completeness, but at the cost of reduced reliability. At high flux densities
the completeness does not appear to have a significant trend with mesh-size, but this is as expected
as small differences in the local noise value are only a tiny fraction of the source total flux density and
therefore unlikely to affect whether a source is detected. 

To further explore why the completeness does not reach unity for the ASKAP simulations, we examined
bright ($S> 1\,$mJy) input sources which were not extracted. The SKADS input list results in many
multiple-component, or extended, sources which the extraction algorithms extract as one source at
the midpoint between the two input sources (Figure \ref{fig:brightexample}). 
These input sources are extracted as a single source with a position further than 0.5 synthesised beam FWHM from the input sources, and hence the input sources are not identified as having been detected. 
In addition to the deblending issue, which accounts for some of the incompleteness, there are also
extended sources with large total flux densities but low peak flux densities
(Figure~\ref{fig:brightexample}). This results in sources being missed by SExtractor as it searches only for pixels above the detection threshold.

\subsection{SFIND}
\label{subsec:sfindcomp}

The reliability and completeness are again estimated for {\sc sfind}. The reliability estimates for
the ASKAP simulation are presented for $\alpha$ values of 0.1, 1, 2, 5 and 10 percent in Table~\ref{tab:rel5}, for an
rmsbox size of 10 beamwidths only. Due to a known bug, apparently triggered by particular
configurations of pixels in a complex source, {\sc sfind} failed to return a source list on this simulation
for the larger rmsbox size (although the background estimation and associated images were
correctly produced). The reliability for all tested values of $\alpha$ ranges from 91.9\% to 92.5\%
for the ASKAP simulation. The reliability is quite flat for this simulation, over typical choices for
$\alpha$. We expect the reliability to be better than 99\% for $\alpha = 0.1$. The values here,
as for SExtractor, are affected by the bright sources in the image which aren't in the original input catalogue. 
For the Hancock et al.\ simulation, and an rmsbox size of 10 beamwidths, the reliability increases from 91\% to 97\% for $\alpha = 10$ to $0.1$ (Table \ref{tab:rel6}). The result is very similar for a
rmsbox size of 20 beamwidths with the Hancock et al.\ simulation.

\begin{figure*}[htb] %  figure placement: here, top, bottom, or page
   \centering
   \includegraphics[width=3.1in]{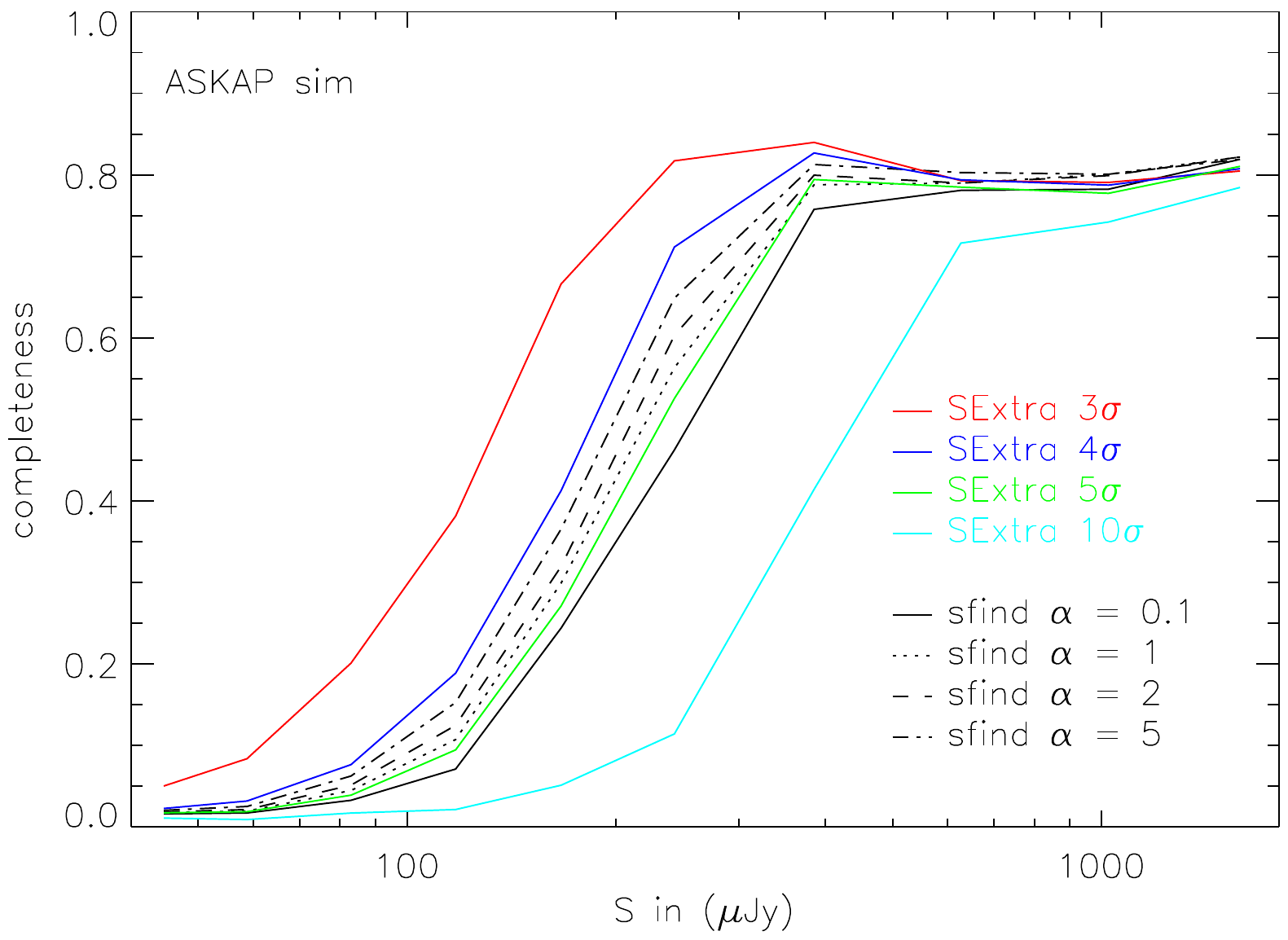} 
    \includegraphics[width=3.1in]{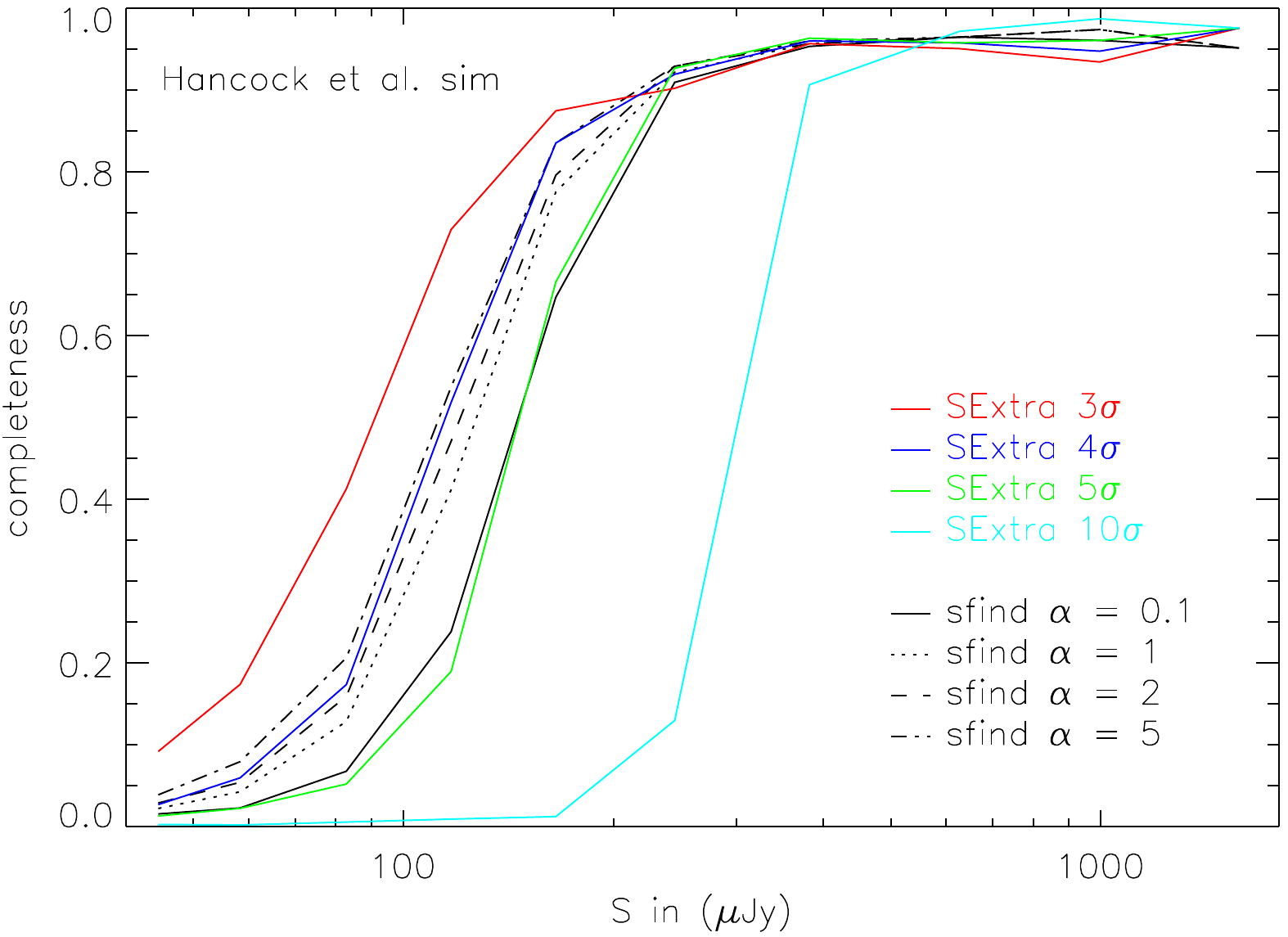} 
   \caption{The completeness as a function of input source flux density for SExtractor and {\sc sfind} on ASKAP (left) and Hancock et al.\ (right) simulations. This is for SExtractor and {\sc sfind} background mesh/rmsbox sizes of 10 beams.}
   \label{fig:comp}
\end{figure*}

\begin{figure*}[htb] %  figure placement: here, top, bottom, or page
   \centering
   \includegraphics[width=2.7in]{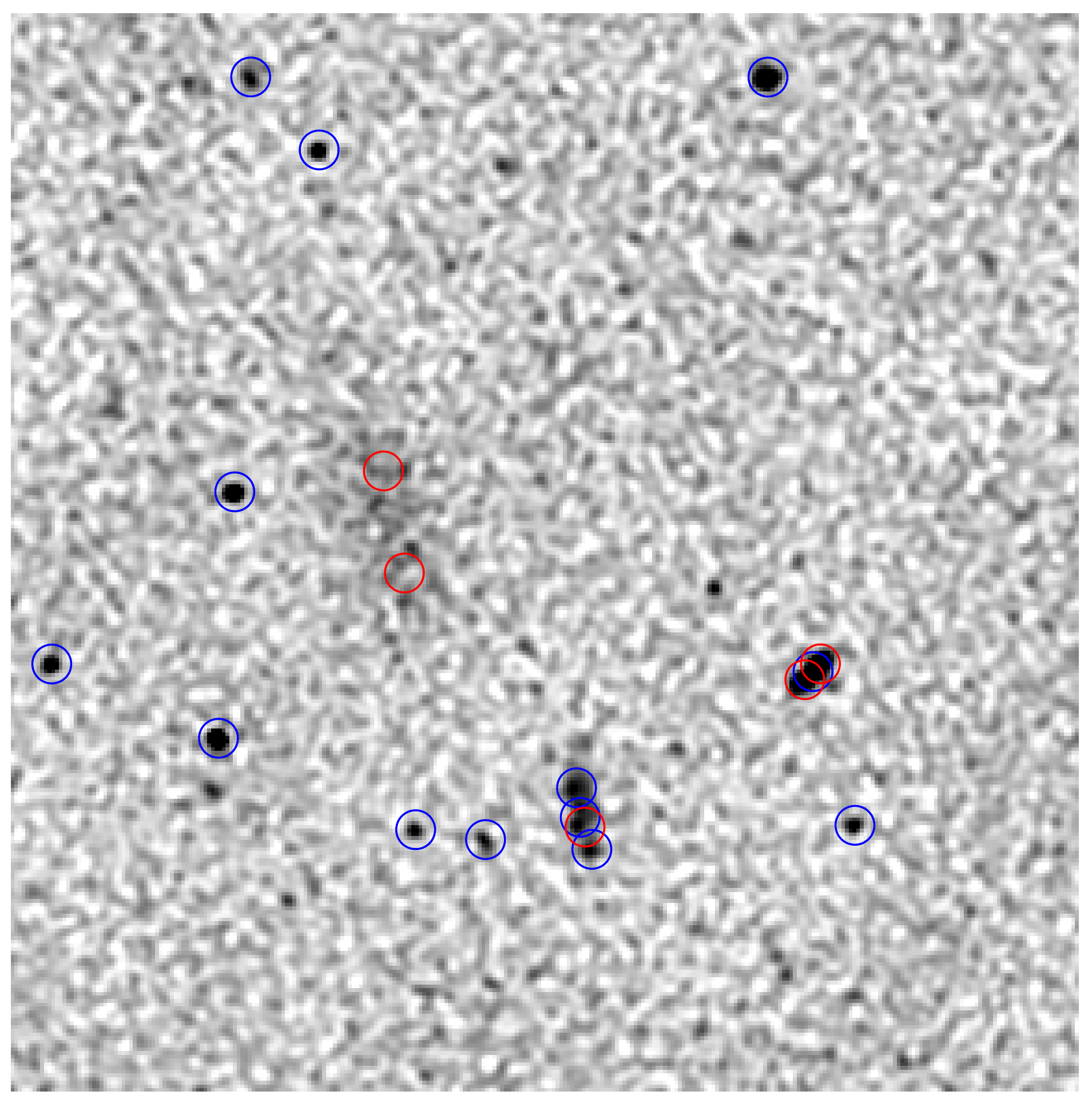} 
   \caption{Examples illustrating why completeness does not reach unity for ASKAP simulations. Red circles are bright ($S> 1\,$mJy) sources in the input list which are not extracted. Blue circles are sources extracted by SExtractor with a threshold of 5$\sigma$. Just left of center, the two red circles are extended sources with large total fluxes but low peak fluxes, and hence they lie below the detection threshold. Right of center is an example of a bright multiple component source which is extracted as one source with a position (blue circle) between the two components (red circles). Bottom of center is another example where poor deblending results in a bright input source that is marked as unextracted due to its poor output position.}
  \label{fig:brightexample}
\end{figure*}

The completeness as a function of input source flux density is plotted in Figure~\ref{fig:comp},
for {\sc sfind} $\alpha$ values of 0.1, 1, 2, and 5 percent.  For the ASKAP simulation, we find the
completeness increases significantly between $100$ and $300\,\mu$Jy for all $\alpha$ values, but
as for SExtractor the completeness does not reach unity at high flux density levels.
Here {\sc sfind} with $\alpha$ values of 1 to 5 gives completeness results that span similar
values to those of SExtractor run with thresholds of $4$ to $5\,\sigma$. The reliability of the
{\sc sfind} sources at these levels of completeness, however, is $\sim 92$\% compared to 80 to 88\% for the SExtractor sources.

In the case of the Hancock et al.\ simulation, an {\sc sfind} $\alpha$ value of 5 results in
completeness similar to SExtractor with a threshold of $4\,\sigma$ (Figure \ref{fig:comp}),
but with slightly worse reliability (93\% compared to 98\%). An {\sc sfind} $\alpha$ value of
0.1\, though, gives a completeness similar to an SExtractor threshold of $5\,\sigma$ as well
as similar reliability (virtually 100\%), but 7\% more sources were extracted by {\sc sfind} with
these parameters. 

\subsection{Selavy}
\label{sec:selavy}

We estimated the reliability and completeness of the Selavy source extraction
on both sets of simulations. The reliability results are shown in Table~\ref{tab:rel3} and \ref{tab:rel4},
for the ASKAP and Hancock et al.\ simulations, respectively. The reliability of Selavy increases
sharply between $3$ and $10\,\sigma$, but the reliability does not reach 100\% for either simulation.
The reliability is 88\% and 85\% for Selavy $20\,\sigma$ runs on the ASKAP and Hancock et al.\
simulations, respectively. It is surprising that the reliability is greater for the ASKAP simulation, as the Hancock et al.\ image has only Gaussian noise and threshold techniques are expected to perform better in the case of pure white noise with no introduced instrumental artefacts.
Although Selavy is more reliable than SExtractor at $3\,\sigma$ for the ASKAP simulation,  it is 5 to 10\% less reliable than SExtractor overall. It also does not reach a reliability of 100\% at the brightest thresholds for the Hancock et al. simulations, whereas this performance is reached by SExtractor. This is likely to be a deblending issue as we found that Selavy split a large proportion of bright sources into separate components, which would then not be matched to an input source. Limiting Selavy to fit only one or two Gaussians per detected ``island" of bright pixels may improve results and this needs to be tested in future work.

\begin{figure*}[htb] %  figure placement: here, top, bottom, or page
   \centering
   \includegraphics[width=3.1in]{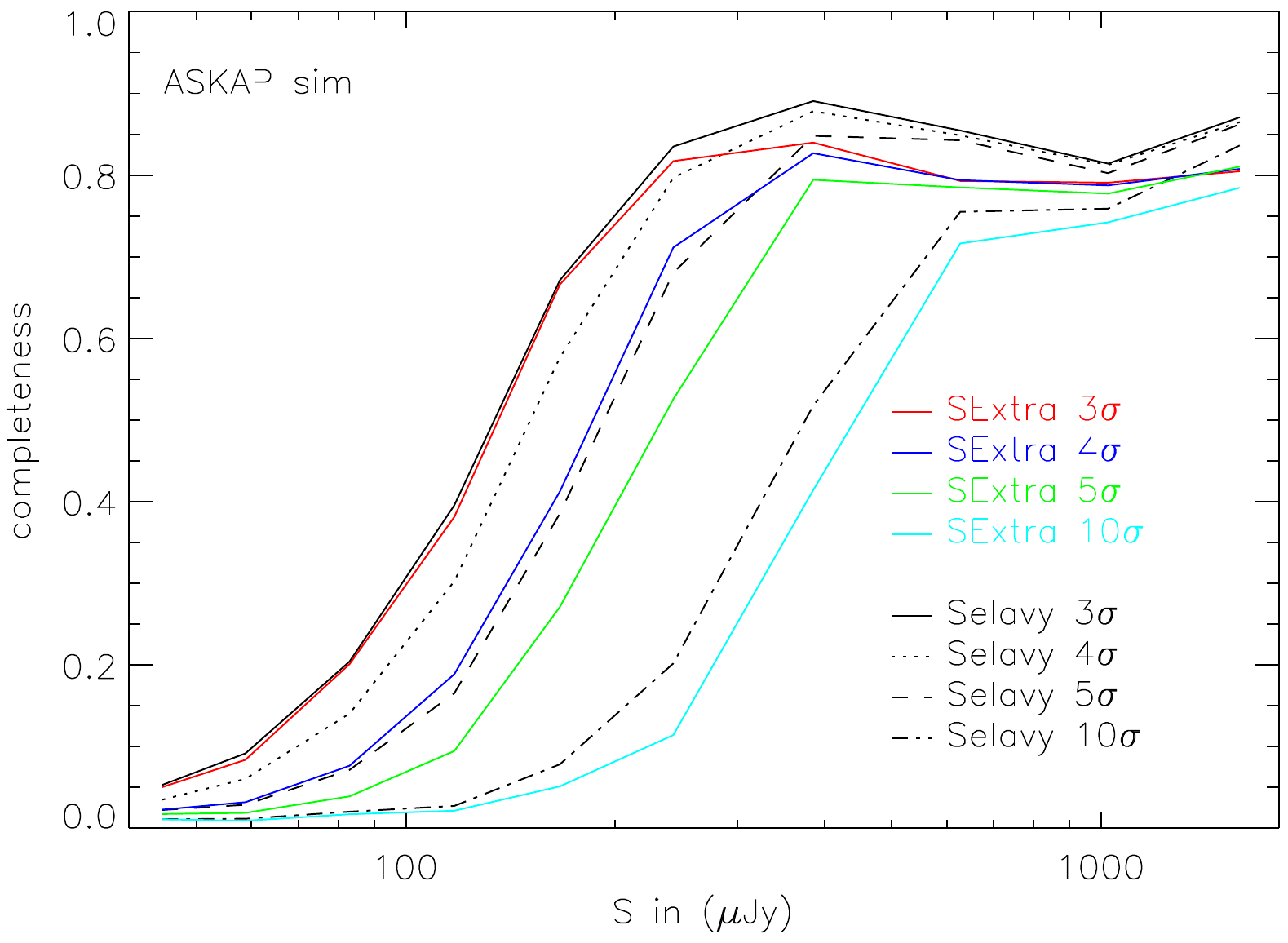} 
    \includegraphics[width=3.1in]{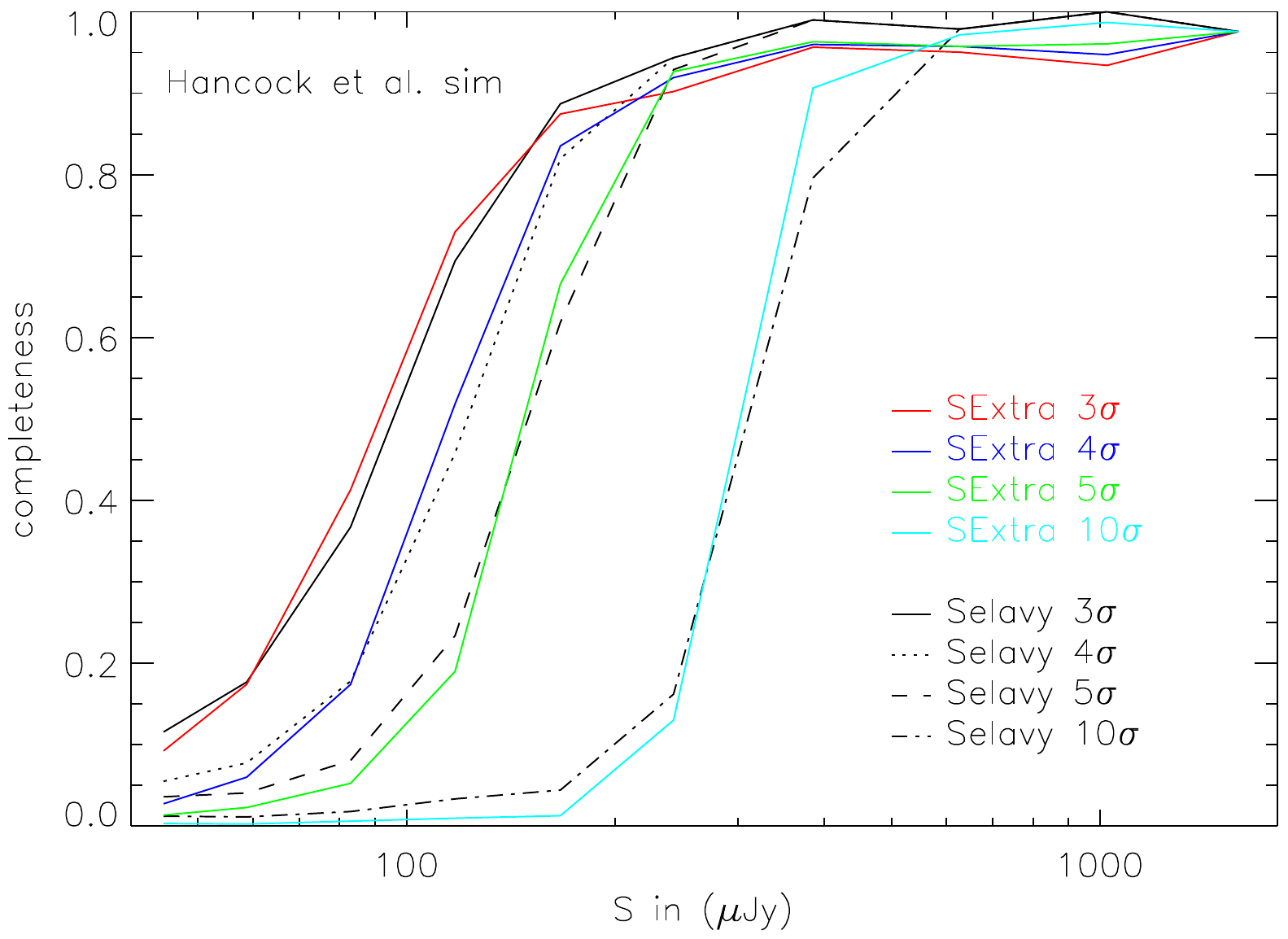} 
   \caption{The completeness as a function of input source flux density for SExtractor and Selavy on ASKAP (left) and Hancock et al.\ (right) simulations. This is for SExtractor and Selavy background mesh/median box sizes of 10 beams.}
   \label{fig:comp2}
\end{figure*}

The completeness of the Selavy algorithm on both sets of simulations is shown in
Figure~\ref{fig:comp2}. The completeness of SExtractor and Selavy are very similar for the Hancock et al. simulation. However, for the ASKAP simulation we find that Selavy is 5 to 10\% more complete than SExtractor for the same nominal threshold limit. For the ASKAP simulations, Selavy reaches a maximum of $\sim$90\% completeness at high flux densities at low thresholds, compared to a maximum completeness of $\sim$80\% with SExtractor. So Selavy appears to be more complete than SExtractor, for the same nominal threshold limit, but this greater completeness comes at the cost of lower reliability.  
 
Comparing Selavy to {\sc sfind}, we find a similar outcome in that Selavy reaches relatively higher completeness but has much lower reliability. In the case of the ASKAP simulations, Selavy with a threshold of 5$\sigma$ has a completeness better than {\sc sfind} with $\alpha$ = 0.1 to 5, however reliability is $\sim$10\% worse. 
For the Hancock et al. simulations the Selavy comparison to {\sc sfind} is much like that with SExtractor, except that Selavy has $\sim$15\% lower reliability than SExtractor. For example, a Selavy threshold of 4$\sigma$ gives a completeness level comparable to {\sc sfind} with $\alpha$ = 2, but with $\sim$30\% less reliability (68\% compared to 95\%). 

 \subsection{Bright Source Region in ASKAP Simulation}
 
\begin{figure*}[htb] %  figure placement: here, top, bottom, or page
   \centering
   \includegraphics[angle=270, width=3.5in]{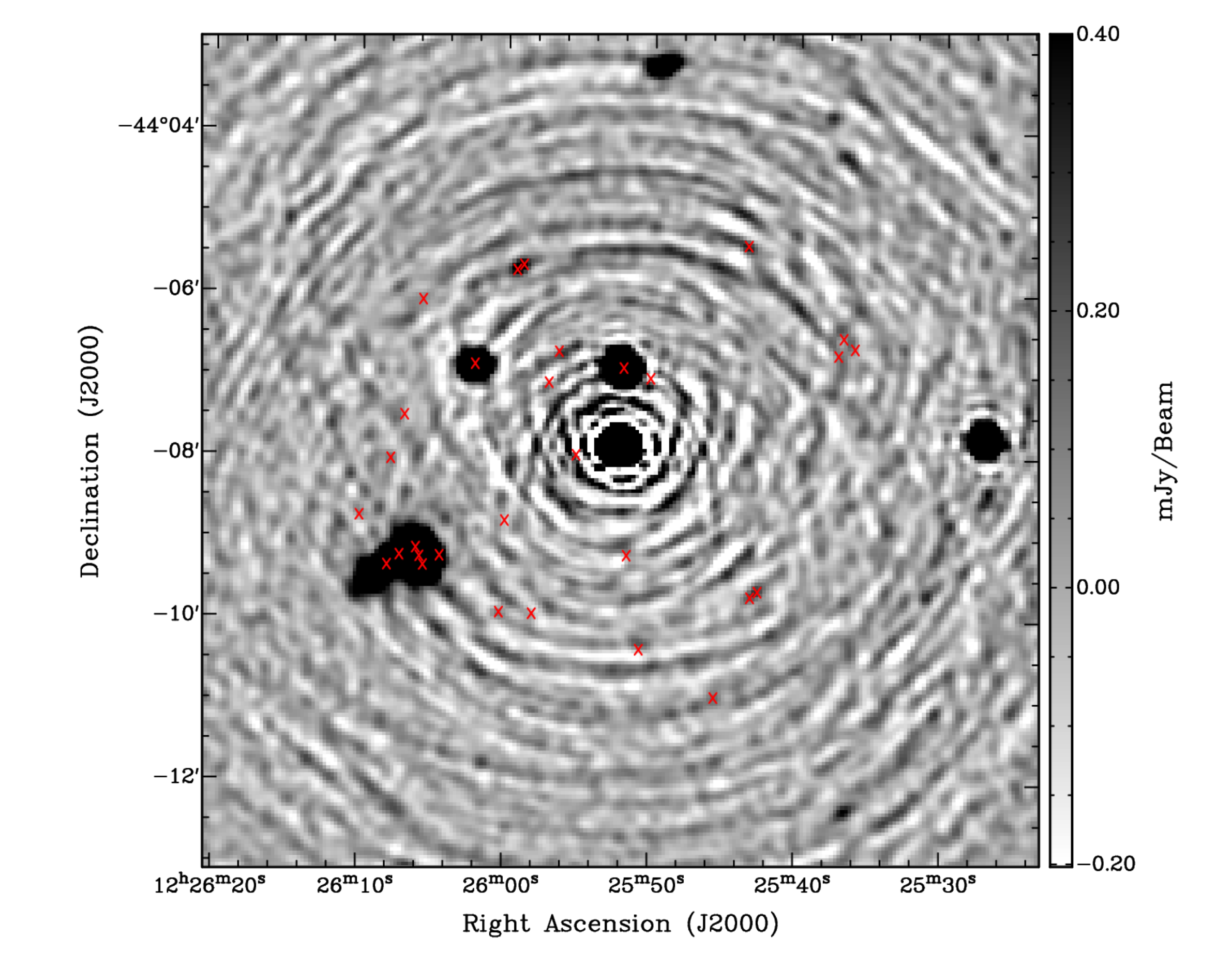} 
   \caption{Region, approximately 30 $\times$ 30 beams, near a bright (3.6 Jy) source in the ASKAP simulation which contains significant sidelobes. Red crosses mark positions of sources in the input catalogue within 20 beams of the bright source.}
   \label{fig:sidelobe}
\end{figure*}

 The ASKAP simulation has several regions near bright sources which contain significant sidelobes. The performance of source extraction algorithms may be degraded in regions with imaging artefacts, so we examine the reliability and completeness of the three source extraction algorithms in one such region (shown in Figure \ref{fig:sidelobe}). This region, in the center of the top-right quadrant of the ASKAP image, has the most significant sidelobes in the ASKAP simulation. 
 The reliability and completeness for two radial distances was investigated: distances less than 10 beams from the bright (3.6 Jy) source, and distances 10 to 20 beams from the bright source. The results are summarised in Tables \ref{tab:relcomp_bright1} to \ref{tab:relcomp_bright3}.  
 
SExtractor is only 2 to 25\% reliable for the region closest to the bright source ($< $10 beams), for the thresholds explored, and different mesh-sizes do not affect the reliability or completeness significantly. {\sc sfind} and Selavy perform better in this region, with reliability reaching 100\% for the most stringent parameters. However, the reliability of Selavy close to the sidelobe-producing source is significantly degraded using a mesh-size of 20 beamwidths. 
 
SExtractor reaches 100\% reliability for high thresholds in the region 10 to 20 beams from the side-lobe producing source, but is only $\sim$10\% reliable at 5 $\sigma$ and  $\sim$6\% reliable at 3$\sigma$.  Mesh-sizes of 20 beamwidths or 10 beamwidths with smoothing give marginally better results than a mesh-size of 10 beamwidths at these low thresholds. {\sc sfind} is 100\% reliable, and Selavy  is 75\% reliable or better, in this region for all the thresholds explored. So while Selavy is less reliable than SExtractor overall (see Section \ref{sec:selavy}), it seems to perform better in this high noise region. The {\sc sfind} generated noise map has values marginally lower than the SExtractor ones in this region, so this confirms the FDR routine is more effective in rejecting false sources compared to a simple peak thresholding technique. 

 The greater noise in this region of the image results in low completeness for all three algorithms. The high reliability of {\sc sfind} and Selavy comes at the cost of lower completeness compared to SExtractor. However, even with a threshold of 3$\sigma$, SExtractor has a maximum completeness of only 26\%. A mesh-size of 10 beamwidths instead of 20 beamwidths results in approximately 5\% worse completeness, at the lowest thresholds, for SExtractor. 
 
 Finally we note that there are only 7 input sources within 10 beams of the bright source, and 23 input sources between a distance of 10 and 20 beams. More analysis on other regions is needed to derive better statistics, but from this work we can conclude that {\sc sfind} and Selavy perform better than SExtractor in regions affected by significant sidelobes from a bright source.

\section{Summary and Conclusions}
\label{conc}

We have tested SExtractor, Selavy and {\sc sfind} to explore the effects of background
and noise estimation, along with two approaches to thresholding, a simple $n\sigma$ level
compared to the false-discovery rate method, on source extraction. 
The tests were performed on two sets of simulations, the ASKAP simulations which are based on SKADS input source catalogue and include instrumental artefacts, and the Hancock et al.\ simulation which has only Gaussian
noise. The Hancock et al.\ simulation is an idealised case that is useful for testing the algorithms in
``perfect" conditions. 

The first step in source extraction is background subtraction and noise estimation. We have confirmed the results from the Planck team (Planck Collaboration 2011) and previous deep continuum radio surveys \citep{huynh2005,schinnerer2007,schinnerer2010} that mesh-sizes of 10 to 20 PSFs or beamwidths produce satisfactory background and noise images. We find that SExtractor background mesh-sizes of 10 and 20 beamwidths produce similar results in terms of reliability. A visual inspection shows the background and noise images are still affected by local bright sources for a mesh-size of 10 beamwidths, but combining a mesh-size of 10 beamwidths with smoothing of 3 meshes produces the highest quality background and noise images. The reliability of the catalogues resulting from 10 and 20 beamwidth mesh-sizes, however, does not differ significantly.

The fact that the background estimation step, so crucial in all the subsequent stages of
source-identification and measurement, still needs to be manually tuned in most existing
source-detection software, is a major concern. The optimum
mesh-size for background subtraction and noise estimation is likely to be image specific,
and to vary perhaps substantially depending on the distribution of sources within the image.
Developing an automated process for setting the mesh-size when implementing the background
and noise properties is clearly a priority, and will need to be developed as part of an automated
pipeline for radio telescopes of the future such as the Australian Square Kilometre Array Pathfinder
(ASKAP) and the Square Kilometre Array (SKA). One possible method could be to use a tree-based
approach, identifying rms noise levels for the whole image and for progressively smaller regions,
so that each pixel can be associated with a ``tree" of rms noise values on each scale. Identifying
the scale for which the rms noise plateaus for each pixel could be a suitable approach, and will
be tested as part of the EMU Design Study.

The thresholding comparison was limited for the ASKAP simulation due to a ceiling in the
completeness values, resulting from bright input sources with small separations that
were not deblended well by any of the algorithms, and by the low peak flux density values for extended sources.
Nevertheless, in this simulation we find that {\sc sfind} with $\alpha$ values of 1 to 5 results in similar completeness to SExtractor run with thresholds of 4 to 5$\sigma$. 
The reliability of the {\sc sfind} sources is higher however, $\sim$92\% for {\sc sfind} compared to 80 -- 88\% for the SExtractor. Selavy results in higher completeness than SExtractor or {\sc sfind} for the ASKAP simulations, but at the cost of lower reliability. In regions with significant artefacts such as sidelobes from bright sources {\sc sfind} and Selavy perform much better than SExtractor in rejecting spurious detections. 

In the case of the Hancock et al.\ simulation, where noise is Gaussian and the sources
more well-separated, SExtractor is 98\% reliable with thresholds of $4\,\sigma$ or greater. 
For the Hancock et al.\ simulation we find that $n\,\sigma$ threshold approach of SExtractor
and the FDR approach of {\sc sfind} perform similarly, although the FDR thresholding
of {\sc sfind} seems to give somewhat better reliability at thresholds that produce
comparable levels of completeness. In this idealised simulation SExtractor and Selavy gives similar completeness but the 
Selavy sources are 15\% to 30\% less reliable.

There is a trade-off between completeness and reliability in source extraction algorithms: parameters which give high completeness result in lower reliability. 
Overall, the false-discovery rate method, as tested with {\sc sfind}, results in more reliable sources than SExtractor or Selavy, for parameters that give similar completeness levels. While more fine-tuning of Selavy, the prototype source finder for EMU, is required, our analysis suggests that the FDR approach is worthwhile pursuing and we recommend this be implemented in Selavy. 

This analysis demonstrates that existing approaches to background and noise estimation
seem to be limited not by the specific algorithms but rather the requirement to select appropriate
mesh-sizes over which to calculate a ``local" background and noise estimate. Future work will
require development of an automated background mesh-size estimation process for the ASKAP
software pipeline, in particular for the EMU images. In addition, a complementary
analysis of the source-fitting and parameter measurement approaches is also underway
(Hancock et al.\ in prep.) which will establish the optimum approaches to these latter stages
of source extraction.

\section*{Acknowledgments} %If needed
We thank the anonymous referee for useful comments that have helped to improved this paper.

%\end{multicols}

%putting in tables here

\begin{table*}[htb] %  figure placement: here, top, bottom, or page
   \centering
   \begin{tabular}{|c|ccccccc|}
   \hline
mesh-size    & \multicolumn{7}{|c|}{Threshold $\sigma$} \\
   \hline
    & 3 & 4 & 5 & 6 & 10 & 15 & 20 \\
3 beamwidths & 43.6\% & 73.7\% & 85.5\% & 88.2\%& 90.7\% & 89.6\% & 89.4\% \\
10 beamwidths & 53.9\% & 80.5\% & 88.2\% & 90.6\% & 92.3\% & 92.7\%& 93.3\% \\
20 beamwidths & 54.4\% & 81.3\% & 88.5\% & 90.1\% & 92.5\% & 92.3\% & 94.0\% \\
10 bwidths + 3$\times$smoothing& 53.9\% & 81.1\% & 88.8\% & 91.3\% & 92.9\% & 93.6\% & 94.4\% \\
%50 beamwidths & 53.5\% & 80.6\% & 88.4\% &  \%& \% & \%& \% \\
100 beamwidths & 54.7\% & 81.7\% & 88.9\% & 91.6\% & 93.1\% & 93.6\%& 93.9\% \\
   \hline
   \end{tabular}   
   \caption{Summary of SExtractor output catalogue reliability for ASKAP simulation.}
   \label{tab:rel1}
\end{table*}

\begin{table*}[htb] %  figure placement: here, top, bottom, or page
   \centering
   \begin{tabular}{|c|ccccccc|}
   \hline
mesh-size    & \multicolumn{7}{|c|}{Threshold $\sigma$} \\
   \hline
    & 3 & 4 & 5 & 6 & 10 & 15 & 20 \\
3 beamwidths & 76.3\% & 91.0\% & 92.8\% & 92.2\% & 89.5\% & 87.5\% & 85.4\% \\
10 beamwidths & 88.4\% & 97.7\% & 98.8\% & 99.0\% & 100.0\% & 100.0\%& 100.0\% \\
20 beamwidths & 89.5\% & 97.7\% & 98.9\% & 99.0\%  & 99.4\% & 100.0\% & 100.0\% \\
10 bwidths + 3$\times$ smoothing& 89.8\% & 97.7\% & 99.1\% & 99.0\% & 99.5\% & 100.0\% & 100.0\% \\
100 beamwidths & 90.6\% & 97.9\% & 99.0\%  & 99.0\%& 99.5\% & 100\%& 100\% \\
   \hline
   \end{tabular}   
   \caption{Summary of SExtractor output catalogue reliability for Hancock et al.\ simulation.}
   \label{tab:rel2}
\end{table*}

\begin{table*}[htb] %  figure placement: here, top, bottom, or page
   \centering
   \begin{tabular}{|c|cccc|}
   \hline
$S$ bin & \multicolumn{4}{|c|}{SExtractor 3$\sigma$}\\
 ($\mu$Jy)  & \multicolumn{4}{|c|}{mesh-size (beamwidths)}  \\
&  3 & 10 & 20 & 100  \\ 
   \hline
40 - 50 & 5.7\%        & 5.0\%     & 4.9\%   & 4.5\%  \\
50 - 70 & 8.9\%        & 8.4\%     & 8.2\%   & 7.5\% \\
70 - 100 & 20.5\%   & 20.1\%   & 19.6\% & 18.2\% \\
100 - 140 & 37.7\% & 38.1\%  & 37.7\% & 35.2\% \\ 
140 - 200 & 64.1\% & 66.6\%  & 66.5\% & 64.5\% \\
200 - 300 & 80.2\% & 81.7\%  & 81.4\% & 81.0\% \\ 
300 - 500 & 83.3\% & 84.0\%  & 83.8\% & 84.2\% \\ 
500 - 800 & 79.7\% & 79.3\%  & 79.9\% & 80.1\% \\
800 - 1300 & 78.1\%& 79.1\% & 78.7\% & 78.6\% \\

   \hline
   \end{tabular}   
   \caption{Summary of completeness for SExtractor $3\sigma$ runs on the ASKAP simulation, for mesh-sizes of various beamwidths as shown.}
   \label{tab:comp}
\end{table*}

\begin{table*}[htb] %  figure placement: here, top, bottom, or page
   \centering
   \begin{tabular}{|c|cccc|}
   \hline
$S$ bin & \multicolumn{4}{|c|}{SExtractor 3$\sigma$}\\
 ($\mu$Jy)  & \multicolumn{4}{|c|}{mesh-size (beamwidths)}  \\
&  3 & 10 & 20 & 100  \\ 
   \hline
40 - 50 & 11.4\%        & 9.2\%     & 8.3\%   & 8.3\%  \\
50 - 70 & 20.7\%        & 17.4\%     & 17.0\%   & 16.6\% \\
70 - 100 & 41.3\%   & 41.3\%   & 41.0\% & 41.0\% \\
100 - 140 & 70.5\% & 73.0\%  & 73.2\% & 72.6\% \\ 
140 - 200 & 84.8\% & 87.4\%  & 87.8\% & 87.6\% \\
200 - 300 & 89.7\% & 90.2\%  & 90.9\% & 90.7\% \\ 
300 - 500 & 94.0\% & 95.7\%  & 95.3\% & 95.3\% \\ 
500 - 800 & 94.3\% & 95.0\%  & 94.3\% & 95.0\% \\
800 - 1300 & 97.4\%& 93.4\% & 93.4\% & 93.4\% \\
   \hline
   \end{tabular}   
   \caption{Summary of completeness for SExtractor $3\sigma$ runs on the Hancock et al.\ simulation, for mesh-sizes of various beamwidths as shown.}
   \label{tab:comp2}
\end{table*}

\begin{table*}[htb] %  figure placement: here, top, bottom, or page
   \centering
   \begin{tabular}{|c|ccccc|}
   \hline
rmsbox-size    & \multicolumn{5}{|c|}{$\alpha$ (\%)} \\
   \hline
& 10  & 5 & 2 & 1 & 0.1 \\ 
10 beamwidths & 91.9 & 92.5\% & 92.3\% & 92.2\% & 92.1\%  \\
   \hline
   \end{tabular}   
   \caption{Summary of {\sc sfind} output catalogue reliability for ASKAP simulation.
   %{\sc sfind gave a bus error for rmsboxes larger than 10 beams.}
   }
   \label{tab:rel5}
\end{table*}

\begin{table*}[htb] %  figure placement: here, top, bottom, or page
   \centering
   \begin{tabular}{|c|ccccc|}
   \hline
rmsbox-size    & \multicolumn{5}{|c|}{$\alpha$ (\%)}\\
   \hline
& 10  & 5 & 2 & 1 & 0.1 \\ 
10 beamwidths & 90.9\% & 93.1\% & 95.2\% & 96.1\% & 97.1\%  \\
20 beamwidths & 92.0\% & 94.5\% & 95.8\% & 96.4\% & 97.3\% \\
   \hline
   \end{tabular}   
   \caption{Summary of {\sc sfind} output catalogue reliability for Hancock et al.\ simulation.}
   \label{tab:rel6}
\end{table*}

\begin{table*}[htb] %  figure placement: here, top, bottom, or page
   \centering
   \begin{tabular}{|c|cccccc|}
   \hline
box-size & \multicolumn{6}{|c|}{Threshold $\sigma$} \\
   \hline
    & 3 & 4 & 5 & 6 & 10 & 20 \\
10 beamwidths & 68.8\% & 76.9\% & 81.3\% & 82.0\% & 85.1\% & 87.8\%  \\
20 beamwidths & 69.6\% & 77.0\% & 80.5\% & 80.4\% & 84.0\% & 86.5\%  \\ \hline
   \end{tabular}   
   \caption{Summary of Selavy output catalogue reliability for ASKAP simulation.}
   \label{tab:rel3}
\end{table*}

\begin{table*}[htb] %  figure placement: here, top, bottom, or page
   \centering
   \begin{tabular}{|c|cccccc|}
   \hline
box-size & \multicolumn{6}{|c|}{Threshold $\sigma$} \\
   \hline
    & 3 & 4 & 5 & 6 & 10 & 20 \\
10 beamwidths & 62.7\% & 68.2\% & 70.6\% & 72.5\% & 77.7\% & 84.6\%  \\
20 beamwidths & 63.4\% & 68.0\% & 70.4\% & 72.4\% & 76.9\% & 84.3\%  \\ \hline
   \end{tabular}   
   \caption{Summary of Selavy output catalogue reliability for Hancock et al. simulation.}
   \label{tab:rel4}
\end{table*}

\begin{table*}[htb] %  figure placement: here, top, bottom, or page
   \centering
   \begin{tabular}{|ccccc|}
   \hline
   \multicolumn{5}{|c|}{Distance $<$ 10 beams} \\ 
mesh-size   & \multicolumn{4}{c|}{Threshold $\sigma$} \\
    & 3 & 5 & 10 & 20 \\
10 beamwidths & 1.9\% (14.3\%)& 2.7\% (14.3\%) & 7.1\% (14.3\%) & 25\%(14.3\%) \\
20 beamwidths & 2.0\% (14.3\%)  & 2.6\% (14.3\%)  & 7.1\% (14.3\%) & 25\%(14.3\%) \\
10 bwidths + 3$\times$smoothing & 2.1\% (14.3\%) & 2.6\% (14.3\%) & 7.1\% (14.3\%)  &  25\%(14.3\%) \\
   \hline
   \multicolumn{5}{|c|}{10 beams $<$ Distance $<$ 20 beams} \\ 
mesh-size   & \multicolumn{4}{c|}{Threshold $\sigma$} \\
    & 3 & 5 & 10 & 20 \\
10 beamwidths & 4.9\% (21.7\%)& 10.3\% (17.4\%) & 100\% (13.0\%) & 100\%(8.7\%) \\
20 beamwidths & 6.0\% (26.1\%)  &11.1\% (17.4\%)  & 100\% (13.0\%) & 100\%(8.7\%) \\
10 bwidths + 3$\times$smoothing & 5.9\% (26.1\%) & 12.5\% (17.4\%) & 100\% (13.0\%)  &  100\%(8.7\%) \\   \hline
   \end{tabular}   
   \caption{Summary of SExtractor output catalogue reliability and completeness (in parentheses) near a bright sidelobe producing source in the ASKAP simulations. Results are for two radial distances: sources less than 10 beams in distance, and sources at 10 to 20 beams in distance.}
   \label{tab:relcomp_bright1}
\end{table*}

\begin{table*}[htb] %  figure placement: here, top, bottom, or page
   \centering
   \begin{tabular}{|ccccc|}
   \hline
   \multicolumn{5}{|c|}{Distance $<$ 10 beams} \\ 
mesh-size   & \multicolumn{4}{c|}{$\alpha$  (\%)}\\
    & 10 & 5 & 1 & 0.1 \\
10 beamwidths & 25\% (14.3\%)& 50\% (14.3\%) & 50\% (14.3\%) & 100\%(14.3\%) \\
   \hline
   \multicolumn{5}{|c|}{10 beams $<$ Distance $<$ 20 beams} \\ 
mesh-size   & \multicolumn{4}{c|}{$\alpha$ (\%)}\\
    & 10 & 5 & 1 & 0.1 \\
10 beamwidths & 100\% (13.0\%)& 100\% (13.0\%) & 100\% (13.0\%) & 100\%(8.7\%) \\ \hline
   \end{tabular}   
   \caption{Summary of {\sc sfind} output catalogue reliability and completeness (in parentheses) near a bright sidelobe producing source in the ASKAP simulations. Results are for two radial distances: sources less than 10 beams in distance, and sources at 10 to 20 beams in distance.}
   \label{tab:relcomp_bright2}
\end{table*}

\begin{table*}[htb] %  figure placement: here, top, bottom, or page
   \centering
   \begin{tabular}{|ccccc|}
   \hline
   \multicolumn{5}{|c|}{Distance $<$ 10 beams} \\ 
mesh-size   & \multicolumn{4}{c|}{Threshold $\sigma$} \\
    & 3 & 5 & 10 & 20 \\
10 beamwidths & 25\% (14.3\%)& 33.3\% (14.3\%) & 100\% (14.3\%) & 100\%(14.3\%) \\
20 beamwidths & 8.3\% (14.3\%)  & 12.5\% (14.3\%)  & 33.3\% (14.3\%) & 50\%(14.3\%) \\
   \hline
   \multicolumn{5}{|c|}{10 beams $<$ Distance $<$ 20 beams} \\ 
mesh-size   & \multicolumn{4}{c|}{Threshold $\sigma$} \\
    & 3 & 5 & 10 & 20 \\
10 beamwidths & 75\% (13.0\%)& 100\% (13.0\%) & 100\% (8.7\%) & 100\%(8.7\%) \\
20 beamwidths & 75\% (13.0\%)  &100\% (13.0\%)  & 100\% (8.7\%) & 100\%(8.7\%) \\ \hline
   \end{tabular}   
   \caption{Summary of Selavy output catalogue reliability and completeness (in parentheses) near a bright sidelobe producing source in the ASKAP simulations. Results are for two radial distances: sources less than 10 beams in distance, and sources at 10 to 20 beams in distance.}
   \label{tab:relcomp_bright3}
\end{table*}

%%% Putting in figures here

\clearpage

\end{document}